\documentclass[11pt]{article}
\usepackage{amsfonts}
\usepackage{amsmath}
\usepackage{amssymb}
\usepackage{indentfirst}
\usepackage{graphicx}
\usepackage[colorlinks]{hyperref}
\numberwithin{equation}{section}
\usepackage{cite}

\begin{document}

\title{Generalized Chern-Simons higher-spin gravity theories\\ in three dimensions}
\author{Ricardo Caroca$^{1}$\thanks{%
rcaroca@ucsc.cl}, \thinspace \thinspace Patrick Concha$^{2}$\thanks{%
patrick.concha@pucv.cl}, Octavio Fierro$^{1}$\thanks{%
ofierro@ucsc.cl} \\
Evelyn Rodr\'{\i}guez$^{3}$\thanks{%
evelyn.rodriguez@edu.uai.cl} \thinspace \thinspace and\thinspace \thinspace
Patricio Salgado-Rebolledo$^{4}$\thanks{%
patricio.salgado@edu.uai.cl} \\
\\
[3pt] {\small {}{}{}$^{1}$}\textit{{\small {}{}{}Departamento de Matem\'{a}%
tica y F\'isica Aplicadas }}{\small {}{}}\\
{\small {}{} \ }\textit{{\small {}{}{}Universidad Cat\'olica de la
Sant\'isima Concepci\'on}}{\small {}{}}\\
{\small {}{} \ {}Alonso de Ribera 2850, Concepci\'on, Chile}\\
[15pt] {\small {}{}{}$^{2}$}\textit{{\small {}{}{}Instituto de F\'isica,
Pontificia Universidad Cat\'olica de Valpara\'iso}}{\small {}{}}\\
{\small {}{} \ {}Casilla 4059, Valpara\'iso, Chile}\\
[15pt] {\small {}{}{}$^{3}$}\textit{{\small {}{}{}Departamento de Ciencias,
Facultad de Artes Liberales}}{\small {}{}}\\
{\small {}{} \ }\textit{{\small {}{}{}Universidad Adolfo Ib\'a\~{n}ez}}%
{\small {}{}}\\
{\small {}{} {}Av. Padre Hurtado 750, Vi\~{n}a del Mar, Chile}\\
[15pt] $^{{\small 4}}$\textit{{\small {}{}{}Facultad de Ingenier\'ia y
Ciencias $\&$ UAI Physics Center}}{\small {}{}}\\
{\small {}{} }\textit{{\small {}{}{}Universidad Adolfo Ib\'a\~{n}ez}}{\small %
{}{}}\\
{\small {}{} {}Avda. Diagonal Las Torres 2640, Pe\~{n}alol\'{e}n, Santiago, Chile}%
\\
}
\maketitle

\begin{abstract}
The coupling of spin-3 gauge fields to three-dimensional Maxwell and $AdS$%
-Lorentz gravity theories is presented. After showing how the usual spin-3 extensions of the $AdS$ and the Poincar\'e algebras in three dimensions can be obtained as expansions of $\mathfrak{sl}%
\left( 3,\mathbb{R}\right)$ algebra, the procedure is generalized so as to define new
higher-spin symmetries. Remarkably, the spin-3 extension of the Maxwell symmetry allows one to introduce a novel gravity
model coupled to higher-spin topological matter with vanishing cosmological constant, which in turn corresponds to  a flat limit of the $AdS$-Lorentz case. We extend our
results to define two different families of higher-spin extensions of three-dimensional Einstein gravity.
\end{abstract}
\bigskip
\bigskip

\section{Introduction}

Higher-spin (HS) fields have received renewed interest over the last years due to their appearance in the spectrum of string theory, as well as in simplified models of the AdS/CFT correspondence \cite{Sundborg, KP, SS0, Sorokin} (for more recent developments see \cite{BJ, GY, GG0, RT, Giombi}). The covariant analysis of free massless HS fieds in four dimensions was described long ago by Fronsdal \cite{Fronsdal}. However, it was realized that in space-times of dimension greater than three, the coupling of HS fields to gravity displays
inconsistencies \cite{Weinberg, AD, BHvNdW, AD2, WW, Bekaert, Porrati, BBS}, particularly due to the non-invariance of the action under HS gauge
transformations. Such obstructions restricts the analysis to gravity coupled to
spin-3/2 fields, which corresponds to supergravity. These no-go theorems were later surpassed by allowing non-minimal couplings and non-local
interactions within the framework of Vasiliev theory, which requires to relax the flat background condition and to introduce an infinite tower of massless HS fields \cite{V1, V2, V3, V4}.

In three dimensions, on the other hand, a consistent coupling of massless HS fields to $AdS$ gravity can be described by means of a Chern-Simons (CS) action whose gauge group is given by two copies of $SL\left( n,\mathbb{R}\right)$\,\cite{Blencowe, BBS2, HR}, where a finite number of interacting HS fields can be considered for finite \textit{n} \cite{ CFPT}. Even though the absence of local degrees of freedom, CS theories are\ known
to possess a rich structure that makes them worth to be
studied. In fact, similarly to what happens in the pure gravity case, the $SL\left( n,\mathbb{R}\right)\times SL\left( n,\mathbb{R}\right)$ CS theory has interesting solutions, such as HS black holes \cite{GK1, PTT2, PTT1, CS1, BHPTT, BCFJ, CIL, GPPTT} and conical singularities \cite{AGGR, CF}. Moreover, the asymptotic symmetry of the theory realizes two copies of the $\mathcal{W}_n$ algebra \cite{CFPT,GGS,CFPT2}, which led to conjecture the duality between three-dimensional HS theories and a $\mathcal{W}_n$ minimal
model CFT in a large-N limit \cite{GG, Ahn}. Remarkably, the no-go results can even be avoided in locally flat three-dimensional spacetimes and the main aforementioned results can be generalized to the case of vanishing cosmological constant \cite{ABFGR, GMPT, GP, GGRR, MPTT, AGPRW}.

In the present paper we further generalize the coupling of spin-3 fields to gravity theories in three dimensions. With the aim of extending the previous results beyond the $AdS$ and the Poincar\'e cases, we address the problem of constructing HS extensions of the Maxwell and $AdS$-Lorentz algebras and their generalizations, given by the $\mathfrak{B}_m$ and the $\mathfrak{C}_m$ algebras. The motivation to do this lies in the fact that these symmetries, have been of recent interest in the context of (super)gravity \cite{BCR, Schrader, Sorokas, AKL, DFIMRSV, SSV, FISV,
HR1, SS, AI, CR2, CRS, CFRS, CIRR, PR}.

Initially, the Maxwell symmetry has been introduced in \cite{BCR, Schrader} to describe a Minkowski spacetime in presence of a constant electromagnetic field background. Later it was presented in \cite{AKL} an alternative geometric scenario to introduce a generalized cosmological constant term. Further generalizations of the Maxwell algebras and their supersymmetric extensions have been then successfully developed by diverse authors in \cite{GKL, GGP, BGKL, KL, AILW, CR1}. Recently, the Maxwell algebra and its generalizations $\mathfrak{B}_m$ have been useful to recover General Relativity from CS and Born-Infeld gravity theories in an appropriate limit \cite{EHTZ, GRCS, CPRS1, CPRS2, CPRS3}. In particular, the $\mathfrak{B}_m$ algebras can be obtained as a flat limit of the so-called $\mathfrak{C}_m$ algebras. The latter have been used to relate diverse (pure) Lovelock gravity theories \cite{CDMR, CDIMR, CMR, CR}. More recently, there has been a particular interest in the three-dimensional Maxwell CS gravity theory \cite{SSV, HR1}. In particular, it has been shown in \cite{CMMRSV} that the gravitational Maxwell field appearing on the Maxwell CS action modifies not only the vacuum of the theory but also its asymptotic sector. Furthermore, it has been pointed out that the asymptotic symmetries of the Maxwell and $AdS$%
-Lorentz gravity theories are given by an enlarged deformation of the $\mathfrak{bms}_{3}$ algebra and three copies of the Virasoro algebra, respectively \cite{CMMRSV, CCRS}, making the study of their HS generalization even more appealing.

A key ingredient in the construction of these algebras is the semigroup expansion method \cite{Sexp, Sexp2, AMNT, ACCSP, ILPR, IKMN}, which combines the structure a given Lie algebra with an abelian semigroup to form a Lie algebra of greater dimension. By applying this method to the $\mathfrak{sl}(3,\mathbb{R})$ algebra we will construct spin-3 extensions of the Maxwell and the $AdS$-Lorentz algebras in three dimensions and the associated CS actions. After studying how the known results of HS gravity in three-dimensions fit in the framework of the expansion method, we extend the three-dimensional Maxwell
\cite{SSV} and $AdS$-Lorentz \cite{Sorokas, DFIMRSV} CS gravity theories to a
more general setup that include spin-3 gauge fields. We also show that Maxwell gravity coupled to spin-3 fields can be recovered
as a flat limit of the HS extension of the $AdS$-Lorentz case. Finally, present two different families of gravity theories coupled to spin-3 fields that correspond to HS extensions of the $\mathfrak{B}_m$ and the $\mathfrak{C}_m$ algebras and generalize the $AdS$ and the $AdS$-Lorentz cases as well as their respective flat
limits. The corresponding CS actions provide novel HS topological matter actions coupled to three-dimensional gravity.

The paper is organized as follows. In the next section we briefly review the $%
SL\left( 3,%
\mathbb{R}
\right) \times SL\left( 3,%
\mathbb{R}
\right) $ CS gravity. Subsequently, in section III, we show that this HS
gravity theory, together with its flat limit, can be alternatively recovered from one copy of $\mathfrak{sl}%
\left( 3,\mathbb{R}\right) $. In
section IV, we construct the HS extension of Maxwell CS gravity as well as the $AdS$%
-Lorentz case, and show that they are related by a flat limit procedure. Section V is devoted to the construction of
HS extensions of the $\mathfrak{B}_{m}$ and $\mathfrak{C}_{m}$ gravities. We conclude
our work with some comments and possible future developments.

\section{Review of $SL\left( 3,%
\mathbb{R}
\right) \times SL\left( 3,%
\mathbb{R}
\right) $ Chern-Simons gravity}

The spin-3 extension of three-dimensional $AdS$ gravity can be formulated as a
CS theory for the group $SL\left( 3,\mathbb{R}
\right) \times SL\left( 3,\mathbb{R}\right) $ \cite{CFPT}. As pure gravity corresponds to the $SL(2,\mathbb{R}%
)\times SL(2,\mathbb{R})$ sector of the theory, the field content of the
full theory is determined by the embedding of the $\mathfrak{sl}(2,\mathbb{R}%
)$ algebra in $\mathfrak{sl}(3,\mathbb{R})$. The $\mathfrak{sl}\left( 3,\mathbb{R}\right) $ algebra is defined by the commutation relations
\begin{eqnarray}
\lbrack L_{i},L_{j}] &=&(i-j)L_{i+j}\,, \\
{}[L_{i},W_{m}] &=&(2i-m)W_{i+m}\,,  \label{eq:sl3} \\
{}[W_{m},W_{n}] &=&\frac{\sigma }{3}(m-n)(2m^{2}+2n^{2}-mn-8)L_{m+n}\,,
\end{eqnarray}%
where $i,j=-1,0,1$ and $m,n=-2,-1,0,1,2$. Here we consider $\sigma <0$\footnote{As explained in \cite{CFPT}, the $\mathfrak{sl}\left( 3,\mathbb{R}\right) $ corresponds to $\sigma <0$ while $\sigma >0$ reproduces the $%
\mathfrak{su}\left( 1,2\right)$
algebra. For completeness, we shall consider an arbitrary $\sigma $, keeping in mind that we are interested in the negative value of $\sigma $.} and the corresponding Killing form in the fundamental representation is normalized such that
\begin{eqnarray}
\langle \ L_{0}L_{0}\;\rangle \  &=&\frac{1}{2}\;,\qquad \qquad \qquad
\qquad \qquad \langle \ W_{0}W_{0}\;\rangle \ \;\;=\;\;-\frac{2}{3}\sigma \,,
\notag \\
\langle \ L_{1}L_{-1}\;\rangle \  &=&-1\;,\;\ \ \ \ \ \ \ \ \ \ \ \ \ \ \ \
\ \ \ \ \ \ \ \ \ \ \langle \ W_{2}W_{-2}\;\rangle \
\;\;=\;\;-4\sigma \,.  \notag \\
\langle \ W_{1}W_{-1}\;\rangle \  &=&\sigma \;,\;\;\qquad   \label{Tr}
\end{eqnarray}%
The algebra $%
\mathfrak{sl}(2,\mathbb{R})$ can be non-trivially embedded in $\mathfrak{sl}%
(3,\mathbb{R})$ in two inequivalent ways: the principal embedding $\left\{
L_{0},L_{\pm 1}\right\} $, which gives rise to an interacting theory of massless spin-$2$ and spin-$3$ fields; and the diagonal embedding $\left\{ \frac{1}{2}%
L_{0},\frac{1}{4}W_{\pm 2}\right\} $, leading to a theory for a spin-$2$ field, two spin-$3/2$ fields and a spin-$1$ current \cite{CS1, AGKP}. As we want to describe gravity coupled
to spin-3 matter fields, the principal embedding of the $\mathfrak{sl}(2,\mathbb{R}%
)$ in $\mathfrak{sl}(3,\mathbb{R})$ will be considered throughout this article. For our purposes it will be convenient to write $\mathfrak{sl}\left( 3,%
\mathbb{R}
\right) $ in the form%
\begin{eqnarray}
\left[ J_{a},J_{b}\right] &=&\epsilon _{abc}J^{c}\,, \label{eq:genJT-1}  \\
\left[ J_{a},T_{bc}\right] &=&\epsilon _{\text{ }a\left( b\right.
}^{m}T_{\left. c\right) m}\,,  \label{eq:genJT-2} \\
\left[ T_{ab},T_{cd}\right] &=&\sigma \left( \eta _{a\left( c\right.
}\epsilon _{\left. d\right) bm}+\eta _{b\left( c\right. }\epsilon _{\left.
d\right) am}\right) J^{m}\,,\label{eq:genJT-3}
\end{eqnarray}%
where the generators $\left\{ J_{a},T_{ab}\right\} $ are related to those of
(\ref{eq:sl3}) by
\begin{equation}\label{eq:genJT}
\begin{array}{cclcclccl}
J_{0} & = & \frac{1}{2}(L_{-1}+L_{1})\,\,,\medskip  & J_{1} & = & \frac{1}{2}
(L_{-1}-L_{1})\,, & J_{2} & = & L_{0}\,, \\
T_{00} & = & \frac{1}{4}(W_{2}+W_{-2}+2W_{0})\,,\medskip  & T_{11} & = &
\frac{1}{4}(W_{2}+W_{-2}-2W_{0})\,, & T_{22} & = & W_{0}\,, \\
T_{01} & = & \frac{1}{4}(W_{2}-W_{-2})\,, & T_{02} & = & \frac{1}{2}
(W_{1}+W_{-1})\,, & T_{12} & = & \frac{1}{2}(W_{1}-W_{-1})\,.
\end{array}
\end{equation}
In this case, instead of the $8$ generators of the fundamental representation, there
are $9$ generators $\left\{ J_{a},T_{ab}=T_{ba}\right\} ;a,b=1,2,3$ ,
plus the constraint $
T_{a}^{a}=0$, were indices are lowered and raised with the metric $\eta _{ab}=diag(-1,1,1)$.

The action of the system is given by
\begin{equation} \label{eq:actionsl3}
S=S_{CS}[A]-S_{CS}[\bar{A}]\,,
\end{equation}%
where $S_{CS}[A]$ corresponds to the CS action
\begin{equation} \label{eq:action}
S_{CS}[A]=\kappa \int \left\langle A\wedge dA+%
\frac{2}{3}A\wedge A\wedge A\right\rangle \,.
\end{equation}
The components of the invariant tensor follow from (\ref{eq:genJT}) and (%
\ref{Tr}), giving
\begin{eqnarray}
\left\langle J_{a}J_{b}\right\rangle &=&\frac{1}{2}\eta _{ab}\,, \label{eq:inv1} \\
\left\langle J_{a}T_{bc}\right\rangle &=&0\,,  \label{eq:inv2} \\
\left\langle T_{ab}T_{bc}\right\rangle &=&-\frac{\sigma }{2}\left( \eta
_{a(c}\eta _{d)b}-\frac{2}{3}\eta _{ab}\eta _{bc}\right) \,,  \label{eq:inv3}
\end{eqnarray}%
and the $\mathfrak{sl}\left( 3,%
\mathbb{R}
\right) $ valued connection one-forms $A$ and $\bar{A}$ have the form
\begin{equation*}
A=\left( \omega ^{a}+\frac{1}{\ell }e^{a}\right) J_{a}+\left( \omega ^{ab}+%
\frac{1}{\ell }e^{ab}\right) T_{ab}\;,\;\;\bar{A}=\left( \omega ^{a}-\frac{1%
}{\ell }e^{a}\right) \bar{J}_{a}+\left( \omega ^{ab}-\frac{1}{\ell }%
e^{ab}\right) \bar{T}_{ab}\;.
\end{equation*}%
The field equations are naturally given by the vanishing of the curvatures
associated to $A$ and $\bar{A}$
\begin{equation*}
dA+A\wedge A=0\;\;\;\;\;d\bar{A}+\bar{A}\wedge \bar{A}=0\,.
\end{equation*}%
As each subset of $\mathfrak{sl}\left( 3,%
\mathbb{R}
\right) $ generators satisfies (\ref{eq:genJT-1}-\ref{eq:genJT-3}) and (\ref{eq:inv1}-\ref{eq:inv3}), the
action (\ref{eq:actionsl3}) takes the form
\begin{eqnarray}
S &=&\frac{k}{2\pi \ell }\int \left[ e^{a}\left( d\omega _{a}+\frac{1}{2}%
\epsilon _{abc}\omega ^{b}\omega ^{c}+2\sigma \epsilon _{aec}\omega
^{cd}\omega _{\ d}^{e}\right) \right.  \notag \\
&&\left. -2\sigma e^{ab}\left( d\omega _{ab}+2\epsilon _{acd}\omega
^{c}\omega _{\ b}^{d}\right) +\frac{1}{6\ell ^{2}}\epsilon
_{abc}e^{a}e^{b}e^{c}+\frac{2\sigma }{\ell ^{2}}\epsilon
_{aec}e^{a}e^{cd}e_{\ d}^{e}\right] \,,\label{Scamp}
\end{eqnarray}%
where, in this case, $\kappa =\frac{k}{4\pi }$ and the CS level $k$ is
related to the Newton constant by
\begin{equation*}
k=\frac{\ell }{4G}\,.
\end{equation*}%
The field equations coming from this action can be expanded around a vacuum
solution and, after using the torsion constraints to express $\omega ^{a}$
and $\omega ^{ab}$ in terms of $e^{a}$ and $e^{ab}$, they reduce to the Fronsdal
equations \cite{Fronsdal} for the space-time metric and a spin-$3$ field.

\section{$SL\left( 3,\mathbb{R}\right) \times SL\left( 3,\mathbb{R}\right) $
gravity from $SL\left( 3,\mathbb{R}\right) $}

In this section, we construct the most general CS action for the gauge group
$SL\left( 3,\mathbb{R}\right) \times SL\left( 3,\mathbb{R}\right) $ by means of the semigroup expansion method \cite{Sexp}, which allows one to obtain the $\mathfrak{sl}\left( 3,\mathbb{R}\right) \times \mathfrak{sl}%
\left( 3,\mathbb{R}\right) $ algebra from $\mathfrak{sl}\left( 3,\mathbb{R}%
\right) $. Furthermore, this method will provide us with the non-vanishing
components of the invariant tensor for the expanded algebra. Note that this kind of construction has already been considered in the context of HS \cite{PV}.

Let us consider the $%
\mathbb{Z}
_{2}=\left\{ \lambda _{0},\lambda _{1}\right\} $ (semi)group, whose elements
satisfy the following multiplication law%
\begin{equation}
\lambda _{\alpha }\lambda _{\beta }=\left\{
\begin{array}{c}
\lambda _{\alpha +\beta },\text{ \ \ \ \ \ \ \ \ \ if }\alpha +\beta \leq 1
\\
\lambda _{\alpha +\beta -2},\text{\ \ \ \ \ \ \ if }\alpha +\beta >1%
\end{array}%
\right. \,.  \label{Z2}
\end{equation}%
Hence, the $%
\mathbb{Z}
_{2}$-expanded algebra is spanned by the set of generators%
\begin{equation}
\mathbb{Z}
_{2}\times \mathfrak{sl}\left( 3,\mathbb{R}%
\right)=\{M_{a},P_{a},M_{ab},P_{ab}\}\,,
\end{equation}%
which can be written in terms of the original ones as follows
\begin{eqnarray}
M_{a} &=&\lambda _{0}J_{a}\,,\text{ \ \ \ \ }\ell P_{a}=\lambda _{1}J_{a}\,,
\\
M_{ab} &=&\lambda _{0}T_{ab}\,,\text{ \ \ }\ell P_{ab}=\lambda _{1}T_{ab}\,.
\end{eqnarray}
Using the semigroup multiplication law (\ref{Z2}) and the commutation
relations of the original algebra (\ref{eq:genJT-1}-\ref{eq:genJT-3}), it
can be shown that the generators of the expanded algebra satisfy the
following commutation relations:
\begin{equation}
\left[ M_{a},M_{b}\right] =\epsilon _{abc}M^{c}\,,\text{ \ \ \ \ }\left[
M_{a},P_{b}\right] =\epsilon _{abc}P^{c}\,,\text{ \ \ \ \ }\left[ P_{a},P_{b}%
\right] =\frac{1}{\ell ^{2}}\epsilon _{abc}M^{c}\,,  \label{2sl3A}
\end{equation}%
\begin{eqnarray}
\left[ M_{a},M_{bc}\right] &=&\epsilon _{\text{ }a\left( b\right.
}^{m}M_{\left. c\right) m}\,,\text{ \ \ \ \ }\left[ M_{a},P_{bc}\right]
=\epsilon _{\text{ }a\left( b\right. }^{m}P_{\left. c\right) m}\,, \\
\left[ P_{a},M_{bc}\right] &=&\epsilon _{\text{ }a\left( b\right.
}^{m}P_{\left. c\right) m}\,,\text{ \ \ \ \ \ }\left[ P_{a},P_{bc}\right] =%
\frac{1}{\ell ^{2}}\epsilon _{\text{ }a\left( b\right. }^{m}M_{\left.
c\right) m}\,,
\end{eqnarray}%
\begin{eqnarray}
\left[ M_{ab},M_{cd}\right] &=&\sigma \left( \eta _{a\left( c\right.
}\epsilon _{\left. d\right) bm}+\eta _{b\left( c\right. }\epsilon _{\left.
d\right) am}\right) M^{m}\,, \\
\left[ M_{ab},P_{cd}\right] &=&\sigma \left( \eta _{a\left( c\right.
}\epsilon _{\left. d\right) bm}+\eta _{b\left( c\right. }\epsilon _{\left.
d\right) am}\right) P^{m}\,, \\
\left[ P_{ab},P_{cd}\right] &=&\frac{\sigma }{\ell ^{2}}\left( \eta
_{a\left( c\right. }\epsilon _{\left. d\right) bm}+\eta _{b\left( c\right.
}\epsilon _{\left. d\right) am}\right) M^{m}\,.  \label{2sl3B}
\end{eqnarray}%
Note that the spin-$2$ generators $\left\{ M_{a},P_{a}\right\} $ (which
satisfy the $\mathfrak{so}\left( 2,2\right) $ algebra) and the spin-$3$
generators $\left\{ M_{ab},P_{ab}\right\} $ satisfy the $\mathfrak{sl}\left(
3,%
\mathbb{R}
\right) \times \mathfrak{sl}\left( 3,%
\mathbb{R}
\right) $ algebra. This can be explicitly seen by redefining the generators in the form
\begin{eqnarray}
M_{a} &=&J_{a}+\bar{J}_{a}\;,\text{ \ \ \ \ \ \ \ }P_{a}\;\;=\;\;\frac{1}{\ell }%
\left( J_{a}-\bar{J}_{a}\right) \;, \\
M_{ab} &=&T_{ab}+\bar{T}_{ab}\;,\;\;\text{ \ \ }P_{ab}\;\;=\;\;\frac{1}{\ell }\left(
T_{ab}-\bar{T}_{ab}\right) \,.
\end{eqnarray}
Remarkably, the semigroup expansion procedure can also be used to
recover the quadratic Casimir of the expanded algebra from the original
Casimir operator. In particular, the $\mathfrak{sl}\left( 3,%
\mathbb{R}
\right) \times \mathfrak{sl}\left( 3,%
\mathbb{R}
\right) $ algebra has the following quadratic Casimir,%
\begin{eqnarray}
C &=&\mu _{0}\left( M_{a}M^{a}+\frac{1}{\ell ^{2}}P_{a}P^{a}-\frac{1}{%
2\sigma }\left[ M_{ab}M^{ab}+\frac{1}{\ell ^{2}}P_{ab}P^{ab}\right] \right)
\notag \\
&&+2\frac{\mu _{1}}{\ell }\left( P_{a}M^{a}-\frac{1}{2\sigma }%
P_{ab}M^{ab}\right) \,,
\end{eqnarray}%
where $\mu _{0}$ and $\mu _{1}$ are arbitrary constants.

In order to write down a CS action for this algebra, we define the one-form
gauge connection,
\begin{equation}
A=\omega ^{a}M_{a}+e^{a}P_{a}+\omega ^{ab}M_{ab}+e^{ab}P_{ab}\text{%
\thinspace }.  \label{one-form}
\end{equation}%
Following Theorem VII.2 of ref.~\cite{Sexp}, the invariant tensor for the $%
\mathfrak{sl}\left( 3,%
\mathbb{R}
\right) \times \mathfrak{sl}\left( 3,%
\mathbb{R}
\right) $ algebra can be obtained from (\ref{eq:inv1}-\ref{eq:inv3}) using the expansion method, which yields
\begin{eqnarray}
\left\langle M_{a}M_{b}\right\rangle &=&\mu _{0}\left\langle
J_{a}J_{b}\right\rangle =\frac{\mu _{0}}{2}\eta _{ab}\,,  \label{it1} \\
\left\langle P_{a}P_{b}\right\rangle &=&\frac{\mu _{0}}{\ell ^{2}}%
\left\langle J_{a}J_{b}\right\rangle =\frac{\mu _{0}}{2\ell ^{2}}\eta
_{ab}\,, \\
\left\langle M_{a}P_{b}\right\rangle &=&\frac{\mu _{1}}{\ell }\left\langle
J_{a}J_{b}\right\rangle =\frac{\mu _{1}}{2\ell }\eta _{ab}\,, \\
\left\langle M_{ab}M_{bc}\right\rangle &=&\mu _{0}\left\langle
T_{ab}T_{bc}\right\rangle =-\frac{\sigma \mu _{0}}{2}\left( \eta _{a(c}\eta
_{d)b}-\frac{2}{3}\eta _{ab}\eta _{dc}\right) \,, \\
\left\langle P_{ab}P_{bc}\right\rangle &=&\frac{\mu _{0}}{\ell ^{2}}%
\left\langle T_{ab}T_{bc}\right\rangle =-\frac{\sigma \mu _{0}}{2\ell ^{2}}%
\left( \eta _{a(c}\eta _{d)b}-\frac{2}{3}\eta _{ab}\eta _{dc}\right) \,, \\
\left\langle M_{ab}P_{bc}\right\rangle &=&\frac{\mu _{1}}{\ell }\left\langle
T_{ab}T_{bc}\right\rangle =-\frac{\sigma \mu _{1}}{2\ell }\left( \eta
_{a(c}\eta _{d)b}-\frac{2}{3}\eta _{ab}\eta _{dc}\right) \,.  \label{it6}
\end{eqnarray}%
Then, the $SL\left( 3,\mathbb{R}\right) \times
SL\left( 3,\mathbb{R}\right) $ CS action (\ref{eq:action}) can be written, modulo
boundary terms, as%
\begin{eqnarray}
S_{CS} &=&\kappa \mu _{0}\int \left[ \frac{1}{2}\left( \omega ^{a}d\omega _{a}+%
\frac{1}{3}\epsilon _{abc}\omega ^{a}\omega ^{b}\omega ^{c}\right) \right. \notag \\
&&\left. +\frac{1}{2\ell ^{2}}e^{a}\left( de_{a}+\epsilon _{abc}\omega
^{b}e^{c}\right) -\sigma \left( \omega _{\text{ }}^{ab}d\omega
_{ab}+2\epsilon _{abc}\omega ^{a}\omega ^{be}\omega _{\text{ }e}^{c}\right) \right.
\notag \\
&&\left. -\frac{\sigma }{\ell ^{2}}e_{\text{ }}^{ab}\left( de_{ab}+2\epsilon
_{acd}\omega ^{c}e_{\text{ }b}^{d}+4\epsilon _{acd}e^{c}\omega _{\text{ }%
b}^{d}\right) \right] \notag \\
&&+\kappa \frac{\mu _{1}}{\ell }\int \left[ e^{a}\left( d\omega _{a}+\frac{1%
}{2}\epsilon _{abc}\omega ^{b}\omega ^{c}-2\sigma \epsilon _{abc}\omega
^{bd}\omega _{\text{ }d}^{c}\right) \right.  \notag \\
&&\left. -2\sigma e^{ab}\left( d\omega _{ab}+2\epsilon _{acd}\omega
^{c}\omega _{\ b}^{d}\right) +\frac{1}{6\ell ^{2}}\epsilon _{abc}\left(
e^{a}e^{b}e^{c}-12\sigma \epsilon _{abc}e^{a}e^{bd}e_{\text{ }d}^{c}\right) %
\right] \,.  \label{sl3sl3}
\end{eqnarray}%
Note that the term proportional to $\mu _{0}$ contains the exotic Lagrangian \cite{Witten}
plus contributions coming from the presence of the spin-3 fields. The term
proportional to $\mu _{1}$, on the other hand, corresponds to the Lagrangian presented in \cite{CFPT}, (\ref{Scamp}), up to a factor 2. Therefore, the above action is the
most general CS action describing the coupling of a spin-3 gauge field to
$AdS$ gravity in three dimensions.

The equations of motion for the gravitational fields are%
\begin{eqnarray}
\mathcal{T}^{a} &\equiv &de^{a}+\epsilon ^{abc}\omega _{b}e_{c}-4\sigma
\epsilon ^{abc}e_{bd}\omega _{c}^{\text{ }d}=0\,,  \label{eom1} \\
\mathcal{R}^{a} &\equiv &d\omega ^{a}+\frac{1}{2}\epsilon ^{abc}\left(
\omega _{b}\omega _{c}+\frac{1}{\ell ^{2}}e_{b}e_{c}\right) -2\sigma
\epsilon ^{abc}\left( \omega _{bd}\omega _{c}^{\text{ }d}+\frac{1}{\ell ^{2}}%
\epsilon ^{abc}e_{bd}e_{c}^{\text{ }d}\right) =0\,,
\end{eqnarray}%
while the field equations of the spin-3 fields read%
\begin{eqnarray}
\mathcal{T}^{ab} &\equiv &de^{ab}+\epsilon ^{cd\left( a\right\vert }\omega
_{c}e_{d}^{\text{ }\left\vert b\right) }+\epsilon ^{cd\left( a\right\vert
}e_{c}\omega _{d}^{\text{ }\left\vert b\right) }=0\,, \\
\mathcal{R}^{ab} &\equiv &d\omega ^{ab}+\epsilon ^{cd\left( a\right\vert
}\omega _{c}\omega _{d}^{\text{ }\left\vert b\right) }+\frac{1}{\ell ^{2}}%
\epsilon ^{cd\left( a\right\vert }e_{c}e_{d}^{\text{ }\left\vert b\right)
}=0\,.  \label{eom4}
\end{eqnarray}%
As shown in ref.~\cite{CFPT}, the
gauge transformations of the theory $\delta A=D\lambda =d\lambda +\left[ A,\lambda \right] \,,$ with gauge parameter%
\begin{equation}
\lambda =\Lambda ^{a}M_{a}+\xi ^{a}P_{a}+\Lambda ^{ab}M_{ab}+\xi
^{ab}P_{ab}\,
\end{equation}%
lead to the following relations for the spin-2 fields:
\begin{eqnarray}
\delta \omega ^{a} &=&D_{\omega }\Lambda ^{a}-\frac{1}{\ell ^{2}}\epsilon
^{abc}\xi _{b}e_{c}\,-4\sigma \epsilon ^{abc}\omega _{bd}\Lambda _{c}^{\text{
}d}-4\frac{\sigma }{\ell ^{2}}\epsilon ^{abc}e_{bd}\xi _{c}^{\text{ }d},
\label{gt1} \\
\delta e^{a} &=&D_{\omega }\xi ^{a}-\epsilon ^{abc}\Lambda _{b}e_{c}-4\sigma
\epsilon ^{abc}\omega _{bd}\xi _{c}^{\text{ }d}-4\sigma \epsilon
^{abc}e_{bd}\Lambda _{c}^{\text{ }d}\,,
\end{eqnarray}%
where, apart from the usual gauge transformations, there are new ones
along the spin-3 gauge parameters $\xi ^{ab}$ and $\Lambda ^{ab}$.
Analogously, the spin-3 fields transform as
\begin{eqnarray}
\delta \omega ^{ab} &=&d\Lambda ^{ab}\,+\epsilon ^{cd\left( a\right\vert
}\omega _{c}\Lambda _{d}^{\text{ }\left\vert b\right) }+\frac{1}{\ell ^{2}}%
\epsilon ^{cd\left( a\right\vert }e_{c}\xi _{d}^{\text{ }\left\vert b\right)
}+\epsilon ^{cd(a}\omega _{\text{ }c}^{b)}\Lambda _{d}+\frac{1}{\ell ^{2}}%
\epsilon ^{cd(a}e_{\text{ }c}^{b)}\xi _{d}\,, \\
\delta e^{ab} &=&d\xi ^{ab}+\epsilon ^{cd\left( a\right\vert }\omega _{c}\xi
_{d}^{\text{ }\left\vert b\right) }+\epsilon ^{cd\left( a\right\vert
}e_{c}\Lambda _{d}^{\text{ }\left\vert b\right) }+\epsilon ^{cd(a}e_{\text{ }%
c}^{b)}\Lambda _{d}+\epsilon ^{cd(a}\omega _{\text{ }c}^{b)}\xi _{d}\,.
\label{gt4}
\end{eqnarray}

\subsection{Poincar\'{e} limit}

It is well known that Poincar\'{e} (super)gravity can be recovered from
the (super) $AdS$ case through a flat limit, which can be generalized to the case of $SL\left( 3,%
\mathbb{R}
\right) \times SL\left( 3,%
\mathbb{R}
\right) $ HS gravity \cite{CFPT, ABFGR, GMPT, GP, MPTT}. Here we present a novel procedure
to obtain the spin-3 extension of Poincar\'{e} algebra through the semigroup expansion
method. In fact, such HS symmetry is obtained by considering an expansion of
the $\mathfrak{sl}\left( 3,%
\mathbb{R}
\right) $ algebra with $S_{E}^{\left( 1\right) }=\left\{ \lambda
_{0},\lambda _{1},\lambda _{2}\right\} $ as the relevant semigroup:%
\begin{equation}
\lambda _{\alpha }\lambda _{\beta }=\left\{
\begin{array}{c}
\lambda _{\alpha +\beta },\text{ \ \ \ \ \ \ \ \ \ if }\alpha +\beta \leq 1
\\
\lambda _{2}\,,\text{\ \ \ \ \ \ \ \ \ \ \ \ \ \ if }\alpha +\beta >1%
\end{array}%
\right.,   \label{SE1}
\end{equation}%
where $\lambda _{2}$ is the zero element of the semigroup. Then, the $%
S_{E}^{\left( 1\right) }$-expanded algebra is spanned by the set of
generators%
\begin{equation}
S_{E}^{\left( 1\right) }\times \mathfrak{sl}\left( 3,\mathbb{R}\right)
=\{M_{a},P_{a},M_{ab},P_{ab}\}\,,
\end{equation}%
where
\begin{eqnarray}
M_{a} &=&\lambda _{0}J_{a}\,,\text{ \ \ \ \ }\ell P_{a}=\lambda _{1}J_{a}\,,
\\
M_{ab} &=&\lambda _{0}T_{ab}\,,\text{ \ \ }\ell P_{ab}=\lambda _{1}T_{ab}\,.
\end{eqnarray}%
Using the semigroup multiplication law (\ref{SE1}) and the commutation
relations of the original algebra (\ref{eq:genJT-1}-\ref{eq:genJT-3}), it
can be shown that the generators of the expanded algebra satisfy:%
\begin{equation}
\left[ M_{a},M_{b}\right] =\epsilon _{abc}M^{c}\,,\text{ \ \ \ \ }\left[
M_{a},P_{b}\right] =\epsilon _{abc}P^{c}\,,\text{ \ \ \ \ }\left[ P_{a},P_{b}%
\right] =0\,,
\end{equation}%
\begin{eqnarray}
\left[ M_{a},M_{bc}\right]  &=&\epsilon _{\text{ }a\left( b\right.
}^{m}M_{\left. c\right) m}\,,\text{ \ \ \ \ }\left[ M_{a},P_{bc}\right]
=\epsilon _{\text{ }a\left( b\right. }^{m}P_{\left. c\right) m}\,, \\
\left[ P_{a},M_{bc}\right]  &=&\epsilon _{\text{ }a\left( b\right.
}^{m}P_{\left. c\right) m}\,,\text{ \ \ \ \ \ }\left[ P_{a},P_{bc}\right]
=0\,,
\end{eqnarray}%
\begin{eqnarray}
\left[ M_{ab},M_{cd}\right]  &=&\sigma \left( \eta _{a\left( c\right.
}\epsilon _{\left. d\right) bm}+\eta _{b\left( c\right. }\epsilon _{\left.
d\right) am}\right) M^{m}\,, \\
\left[ M_{ab},P_{cd}\right]  &=&\sigma \left( \eta _{a\left( c\right.
}\epsilon _{\left. d\right) bm}+\eta _{b\left( c\right. }\epsilon _{\left.
d\right) am}\right) P^{m}\,, \\
\left[ P_{ab},P_{cd}\right]  &=&0\,,
\end{eqnarray}%
which corresponds to the spin-3 extension of the Poincar\'{e} algebra. As it is known, this symmetry
can also be recovered as a flat limit $\ell \rightarrow \infty $ of the $%
\mathfrak{sl}\left( 3,%
\mathbb{R}
\right) \times \mathfrak{sl}\left( 3,%
\mathbb{R}
\right) $ algebra. However, as pointed out in ref.~\cite{CFR}, in order to
apply the flat limit at the level of the action (\ref{sl3sl3}) it is
necessary to consider an appropriate redefinition of the constant appearing
in the invariant tensor (\ref{it1}-\ref{it6}),%
\begin{equation}
\mu _{1}\rightarrow \ell \mu _{1}\,.  \label{resc}
\end{equation}%
This redefinition\ is also required to obtain appropriately the quadratic
Casimir of this algebra in the flat limit%
\begin{equation}
C=\mu _{0}\left( M_{a}M^{a}-\frac{1}{2\sigma }M_{ab}M^{ab}\right) +2\mu
_{1}\left( P_{a}M^{a}-\frac{1}{2\sigma }P_{ab}M^{ab}\right) \,.
\end{equation}%
Note that the invariant tensor
for the Poincar\'{e} gravity coupled to spin-3 fields is recovered in this
limit%
\begin{eqnarray}
\left\langle M_{a}M_{b}\right\rangle &=&\frac{\mu _{0}}{2}\eta _{ab}\,,\text{
\ \ }\left\langle M_{a}P_{b}\right\rangle =\frac{\mu _{1}}{2}\eta _{ab} \,,\\
\left\langle M_{ab}M_{bc}\right\rangle &=&-\mu _{0}\frac{\sigma }{2}\left(
\eta _{a(c}\eta _{d)b}-\frac{2}{3}\eta _{ab}\eta _{dc}\right) \,, \\
\left\langle M_{ab}P_{bc}\right\rangle &=&-\mu _{1}\frac{\sigma }{2}\left(
\eta _{a(c}\eta _{d)b}-\frac{2}{3}\eta _{ab}\eta _{dc}\right) \,.
\end{eqnarray}%
It is important to note that this invariant tensor can alternatively be obtained by means of the semigroup expansion procedure.
In the same way, the corresponding CS action can then be either constructed by considering the connection one-form (\ref{one-form}) or taking the limit $\ell \rightarrow \infty $ in (\ref{sl3sl3}) after implementing (\ref{resc}). This leads to
\begin{eqnarray}
S &=&\kappa \mu _{0}\int \left[ \frac{1}{2}\left( \omega ^{a}d\omega
_{a}+\frac{1}{3}\epsilon _{abc}\omega ^{a}\omega ^{b}\omega ^{c}\right)
-\sigma \left( \omega _{\text{ }}^{ab}d\omega _{ab}+2\epsilon _{abc}\omega
^{a}\omega ^{be}\omega _{\text{ }e}^{c}\right) \right]  \notag \\
&&+\kappa \mu _{1}\int \left[ e^{a}\left( d\omega _{a}+\frac{1}{2}\epsilon
_{abc}\omega ^{b}\omega ^{c}-2\sigma \epsilon _{abc}\omega ^{bd}\omega _{%
\text{ }d}^{c}\right) \right.  \notag \\
&&\left. -2\sigma e^{ab}\left( d\omega _{ab}+2\epsilon _{acd}\omega
^{c}\omega _{\ b}^{d}\right) \right] \,.  \label{poinc3}
\end{eqnarray}%
The CS action (\ref{poinc3}) describes the most general coupling of spin-3
fields to the Poincar\'{e} gravity. Here an exotic term is present besides
the usual Poincar\'{e} CS action coupled to spin-3 fields introduced in \cite%
{ABFGR, GMPT, GP, MPTT}. Interestingly, the flat limit can also be applied to the
equations of motion and the gauge transformations. In fact, considering $%
\ell \rightarrow \infty $ in (\ref{eom1}-\ref{eom4}), we obtain
\begin{eqnarray}
\mathcal{T}^{a} &\equiv &de^{a}+\epsilon ^{abc}\omega _{b}e_{c}-4\sigma
\epsilon ^{abc}e_{bd}\omega _{c}^{\text{ }d}=0\,, \\
\mathcal{R}^{a} &\equiv &d\omega ^{a}+\frac{1}{2}\epsilon ^{abc}\omega
_{b}\omega _{c}-2\sigma \epsilon ^{abc}\omega _{bd}\omega _{c}^{\text{ }%
d}=0\,,
\end{eqnarray}%
\begin{eqnarray}
\mathcal{T}^{ab} &\equiv &de^{ab}+\epsilon ^{cd\left( a\right\vert }\omega
_{c}e_{d}^{\text{ }\left\vert b\right) }+\epsilon ^{cd\left( a\right\vert
}e_{c}\omega _{d}^{\text{ }\left\vert b\right) }=0\,, \\
\mathcal{R}^{ab} &\equiv &d\omega ^{ab}+\epsilon ^{cd\left( a\right\vert
}\omega _{c}\omega _{d}^{\text{ }\left\vert b\right) }=0\,,
\end{eqnarray}%
and in (\ref{gt1}-\ref{gt4}) we find%
\begin{eqnarray}
\delta \omega ^{a} &=&D_{\omega }\Lambda ^{a}\,-4\sigma \epsilon
^{abc}\omega _{bd}\Lambda _{c}^{\text{ }d}, \\
\delta e^{a} &=&D_{\omega }\xi ^{a}-\epsilon ^{abc}\Lambda _{b}e_{c}-4\sigma
\epsilon ^{abc}\omega _{bd}\xi _{c}^{\text{ }d}-4\sigma \epsilon
^{abc}e_{bd}\Lambda _{c}^{\text{ }d}\,,
\end{eqnarray}%
\begin{eqnarray}
\delta \omega ^{ab} &=&d\Lambda ^{ab}\,+\epsilon ^{cd\left( a\right\vert
}\omega _{c}\Lambda _{d}^{\text{ }\left\vert b\right) }+\epsilon
^{cd(a}\omega _{\text{ }c}^{b)}\Lambda _{d}, \\
\delta e^{ab} &=&d\xi ^{ab}+\epsilon ^{cd\left( a\right\vert }\omega _{c}\xi
_{d}^{\text{ }\left\vert b\right) }+\epsilon ^{cd\left( a\right\vert
}e_{c}\Lambda _{d}^{\text{ }\left\vert b\right) }+\epsilon ^{cd(a}e_{\text{ }%
c}^{b)}\Lambda _{d}+\epsilon ^{cd(a}\omega _{\text{ }c}^{b)}\xi _{d}\,.
\end{eqnarray}

\section{Coupling spin-3 fields to extended CS gravity}

One way to generalize three-dimensional CS gravity theories consists in
introducing additional fields into the gauge connection. The minimal
extension can be constructed by adding an extra generator to the former
space-time symmetry. Of recent interest are the Maxwell and $AdS$-Lorentz
algebras, which have led to interesting results. In particular, diverse
gravity theories can be recovered from CS and Born-Infeld gravity
models based on those symmetries \cite{EHTZ, GRCS, CPRS1, CPRS2, CPRS3,
CDMR, CDIMR, CMR, CR}.

In this section, using the semigroup expansion mechanism \cite{Sexp}, we present the
coupling of spin-3 fields to the three-dimensional CS gravity based on the
Maxwell and $AdS$-Lorentz symmetries. In particular, the Maxwell gravity
coupled to HS fields can be alternatively obtained as a flat limit of the HS
$AdS$-Lorentz gravity.

\subsection{Maxwell gravity coupled to spin-3 fields}

In this section, we will describe the coupling of spin-3 gauge fields to
Maxwell gravity in three dimensions. For this purpose we will consider $%
S_{E}^{\left( 2\right) }=\left\{ \lambda _{0},\lambda _{1},\lambda
_{2},\lambda _{3}\right\} $ as the abelian semigroup, whose elements satisfy
the following multiplication law
\begin{equation}
\begin{tabular}{l|llll}
$\lambda _{3}$ & $\lambda _{3}$ & $\lambda _{3}$ & $\lambda _{3}$ & $\lambda
_{3}$ \\
$\lambda _{2}$ & $\lambda _{2}$ & $\lambda _{3}$ & $\lambda _{3}$ & $\lambda
_{3}$ \\
$\lambda _{1}$ & $\lambda _{1}$ & $\lambda _{2}$ & $\lambda _{3}$ & $\lambda
_{3}$ \\
$\lambda _{0}$ & $\lambda _{0}$ & $\lambda _{1}$ & $\lambda _{2}$ & $\lambda
_{3}$ \\ \hline
& $\lambda _{0}$ & $\lambda _{1}$ & $\lambda _{2}$ & $\lambda _{3}$%
\end{tabular}
\label{sml}
\end{equation}%
and $\lambda _{3}=0_{S}$ represents the zero element of the semigroup. Hence, the $S_{E}^{\left( 2\right) }$-expanded
algebra is spanned by the set of generators

\begin{equation}
S_{E}^{\left( 2\right) }\times \mathfrak{sl}(3,\mathbb{R})=\left\{
M_{a},P_{a},Z_{a},M_{ab},P_{ab},Z_{ab}\right\} \,,
\end{equation}%
where
\begin{equation*}
\begin{tabular}{lll}
$M_{a}=\lambda _{0}J_{a}\,\,,$ & $\ell P_{a}=\lambda _{1}J_{a}\,,$ & $\text{%
\ }\ell ^{2}Z_{a}=\lambda _{2}J_{a}\,,$ \\
$M_{ab}=\lambda _{0}T_{ab}\,,$ & $\ell P_{ab}=\lambda _{1}T_{ab}\,,$ & $\ell
^{2}Z_{ab}=\lambda _{2}T_{ab}\,.$%
\end{tabular}%
\end{equation*}
Using the semigroup multiplication law (\ref{sml}) and the commutation
relations of the original algebra $\mathfrak{sl}(3,\mathbb{R})$, it can be shown that
the generators of the expanded algebra satisfy the following commutation
relations:
\begin{eqnarray}
\left[ M_{a},M_{b}\right]  &=&\epsilon _{abc}M^{c}\,,\text{ \ \ \ \ }\left[
M_{a},P_{b}\right] =\epsilon _{abc}P^{c}\,,\text{ \ \ \ \ }\left[ P_{a},P_{b}%
\right] =\epsilon _{abc}Z^{c}\,,\, \\
\left[ M_{a},Z_{b}\right]  &=&\epsilon _{abc}Z^{c}\,\,,
\end{eqnarray}%
\begin{eqnarray}
\left[ M_{a},M_{bc}\right]  &=&\epsilon _{\text{ \ }a\left( b\right.
}^{m}M_{\left. c\right) m}\,,\text{ \ \ \ \ }\left[ M_{a},P_{bc}\right]
\;\;=\;\;\epsilon _{\text{ \ }a\left( b\right. }^{m}P_{\left. c\right) m}\,, \\
\left[ P_{a},M_{bc}\right]  &=&\epsilon _{\text{ \ }a\left( b\right.
}^{m}P_{\left. c\right) m}\,,\text{ \ \ \ \ \ \ }\left[ P_{a},P_{bc}\right]
\;\;=\;\;\epsilon _{\text{ \ }a\left( b\right. }^{m}Z_{\left. c\right) m}\,, \\
\left[ Z_{a},M_{bc}\right]  &=&\epsilon _{\text{ \ }a\left( b\right.
}^{m}Z_{\left. c\right) m}\,,\text{ \ \ \ \ \ }\left[ M_{a},Z_{bc}\right]
\;\;=\;\;\epsilon _{\text{ \ }a\left( b\right. }^{m}Z_{\left. c\right) m}\,\,,
\end{eqnarray}%
\begin{eqnarray}
\left[ M_{ab},M_{cd}\right]  &=&\sigma \left( \eta _{a\left( c\right.
}\epsilon _{\left. d\right) bm}+\eta _{b\left( c\right. }\epsilon _{\left.
d\right) am}\right) M^{m}\,, \\
\left[ M_{ab},P_{cd}\right]  &=&\sigma \left( \eta _{a\left( c\right.
}\epsilon _{\left. d\right) bm}+\eta _{b\left( c\right. }\epsilon _{\left.
d\right) am}\right) P^{m}\,, \\
\left[ M_{ab},Z_{cd}\right]  &=&\sigma \left( \eta _{a\left( c\right.
}\epsilon _{\left. d\right) bm}+\eta _{b\left( c\right. }\epsilon _{\left.
d\right) am}\right) Z^{m}\,, \\
\left[ P_{ab},P_{cd}\right]  &=&\sigma \left( \eta _{a\left( c\right.
}\epsilon _{\left. d\right) bm}+\eta _{b\left( c\right. }\epsilon _{\left.
d\right) am}\right) Z^{m}\,, \\
\text{others} &=&0\,.
\end{eqnarray}%
This algebra describes the coupling of the spin-3 generators $\left\{
M_{ab},P_{ab},Z_{ab}\right\} $ to the Maxwell symmetry (also known in the literature as $\mathfrak{B}%
_{4}$) generated by $\left\{
M_{a},P_{a},Z_{a}\right\} $. Note that, similarly to what happens with the translation generators in the pure gravity case, the generators $P_{ab}$ are
no longer abelian due to the presence of the new generators $Z_{a}$ and $Z_{ab}$ . The
introduction of the additional spin-3 generator $Z_{ab}$ is therefore required in order to consistently couple
spin-3 fields to the Maxwell algebra. The new HS algebra obtained here has the quadratic
Casimir%
\begin{eqnarray}
C &=&\mu _{0}\left( M_{a}M^{a}-\frac{1}{2\sigma }M_{ab}M^{ab}\right) +2\mu
_{1}\left( P_{a}M^{a}-\frac{1}{2\sigma }P_{ab}M^{ab}\right)   \notag \\
&&+\mu _{2}\left[ P_{a}P^{a}+2Z_{a}M^{a}-\frac{1}{2\sigma }\left(
P_{ab}P^{ab}+2Z_{ab}M^{ab}\right) \right] \,,  \label{invmax}
\end{eqnarray}%
allowing us to define a non-degenerate bilinear form.

In order to write down the CS action invariant under the HS extension of the Maxwell algebra, we consider the one-form gauge connection%
\begin{equation}
A=\omega ^{a}M_{a}+e^{a}P_{a}+k^{a}Z_{a}+\omega
^{ab}M_{ab}+e^{ab}P_{ab}+k^{ab}Z_{ab}\text{ .}  \label{max one-form}
\end{equation}%
and the corresponding invariant tensor%
\begin{eqnarray}
\left\langle M_{a}M_{b}\right\rangle  &=&\mu _{0}\left\langle
J_{a}J_{b}\right\rangle _{\mathfrak{sl}(3,%
\mathbb{R}
)}=\frac{\mu _{0}}{2}\eta _{ab}\,, \\
\left\langle P_{a}P_{b}\right\rangle  &=&\frac{\tilde{\mu}_{2}}{\ell ^{2}}%
\left\langle J_{a}J_{b}\right\rangle _{\mathfrak{sl}(3,%
\mathbb{R}
)}=\frac{\mu _{2}}{2}\eta _{ab}\,, \\
\left\langle M_{a}P_{b}\right\rangle  &=&\frac{\tilde{\mu}_{1}}{\ell }%
\left\langle J_{a}J_{b}\right\rangle _{\mathfrak{sl}(3,%
\mathbb{R}
)}=\frac{\mu _{1}}{2}\eta _{ab}\,, \\
\left\langle M_{a}Z_{b}\right\rangle  &=&\frac{\tilde{\mu}_{2}}{\ell ^{2}}%
\left\langle J_{a}J_{b}\right\rangle _{\mathfrak{sl}(3,%
\mathbb{R}
)}=\frac{\mu _{2}}{2}\eta _{ab}\,,
\end{eqnarray}%
\begin{eqnarray}
\left\langle M_{ab}M_{bc}\right\rangle  &=&\mu _{0}\left\langle
T_{ab}T_{bc}\right\rangle _{\mathfrak{sl}(3,%
\mathbb{R}
)}=-\mu _{0}\frac{\sigma }{2}\left( \eta _{a(c}\eta _{d)b}-\frac{2}{3}\eta
_{ab}\eta _{dc}\right) \,, \\
\left\langle P_{ab}P_{bc}\right\rangle  &=&\frac{\tilde{\mu}_{2}}{\ell ^{2}}%
\left\langle T_{ab}T_{bc}\right\rangle _{\mathfrak{sl}(3,%
\mathbb{R}
)}=-\mu _{2}\frac{\sigma }{2}\left( \eta _{a(c}\eta _{d)b}-\frac{2}{3}\eta
_{ab}\eta _{dc}\right) \,, \\
\left\langle M_{ab}P_{bc}\right\rangle  &=&\frac{\tilde{\mu}_{1}}{\ell }%
\left\langle T_{ab}T_{bc}\right\rangle _{\mathfrak{sl}(3,%
\mathbb{R}
)}=-\mu _{1}\frac{\sigma }{2}\left( \eta _{a(c}\eta _{d)b}-\frac{2}{3}\eta
_{ab}\eta _{dc}\right) \,, \\
\left\langle M_{ab}Z_{bc}\right\rangle  &=&\frac{\tilde{\mu}_{2}}{\ell ^{2}}%
\left\langle T_{ab}T_{bc}\right\rangle _{\mathfrak{sl}(3,%
\mathbb{R}
)}=-\mu _{2}\frac{\sigma }{2}\left( \eta _{a(c}\eta _{d)b}-\frac{2}{3}\eta
_{ab}\eta _{dc}\right) \,,
\end{eqnarray}%
where we have defined $\mu _{1}=\tilde{\mu}_{1}/\ell ~$and $\mu _{2}=\tilde{%
\mu}_{2}/\ell ^{2}$. Then, considering the invariant tensor and the one-form
connection (\ref{max one-form}) the CS action reads%
\begin{eqnarray}
S &=&\kappa\int \mu _{0}\left[ \frac{1}{2}\left( \omega ^{a}d\omega
_{a}+\frac{1}{3}\epsilon _{abc}\omega ^{a}\omega ^{b}\omega ^{c}\right)
-\sigma \left( \omega _{\text{ }b}^{a}d\omega _{\text{ }a}^{b}+2\epsilon
_{abc}\omega ^{a}\omega ^{bd}\omega _{\text{ }d}^{c}\right) \right]   \notag
\\
&&+\mu _{1}\left[ e^{a}\left( d\omega _{a}+\frac{1}{2}\epsilon _{abc}\omega
^{b}\omega ^{c}-2\sigma \epsilon _{abc}\omega ^{bd}\omega _{\text{ }%
d}^{c}\right) -2\sigma e^{ab}\left( d\omega _{ab}+2\epsilon _{acd}\omega
^{c}\omega _{\ b}^{d}\right) \right]   \notag \\
&&+\mu _{2}\left[ \frac{1}{2}e^{a}\left( de_{a}+\epsilon _{abc}\omega
^{b}e^{c}\right) +k^{a}\left( d\omega _{a}+\frac{1}{2}\epsilon _{abc}\omega
^{b}\omega ^{c}\right) -\sigma e_{\text{ }}^{ab}\left( de_{ab}+2\epsilon
_{acd}\omega ^{c}e_{\text{ }b}^{d}+4\epsilon _{acd}e^{c}\omega _{\text{ }%
b}^{d}\right) \right.   \notag \\
&&\left. -2\sigma \left( \omega ^{ab}dk_{ab}+\epsilon _{abc}k^{a}\omega
^{be}\omega _{\text{ }e}^{c}+2\epsilon _{abc}\omega ^{a}k^{be}\omega _{\text{
}e}^{c}\right) \right] \,.  \label{Maxspin3}
\end{eqnarray}%
This action describes the coupling of spin-3 gauge fields to CS Maxwell
gravity, which splits in three different sectors
proportional to $\mu _{0},\mu _{1}$ and $\mu _{2}$. One can see that the
term proportional to $\mu _{1}$ corresponds to an Euler type CS form while the
term proportional to $\mu _{0}$ and $\mu _{2}$ are Pontryagin type CS forms.
Interestingly, as in the Poincar\'{e} case, the action (\ref{Maxspin3}) does
not contain the cosmological term. The equations of motion for the
spin-2 fields are given by%
\begin{eqnarray}
\mathcal{T}^{a} &\equiv &de^{a}+\epsilon ^{abc}\omega _{b}e_{c}-4\sigma
\epsilon ^{abc}e^{bd}\omega _{c}^{\text{ }d}=0\,, \label{hsm1}\\
\mathcal{R}^{a} &\equiv &d\omega ^{a}+\frac{1}{2}\epsilon ^{abc}\omega
_{b}\omega _{c}-2\sigma \epsilon ^{abc}\omega _{bd}\omega _{c}^{\text{ }%
d}=0\,, \\
\mathcal{F}^{a} &\equiv &dk^{a}+\epsilon ^{abc}\omega _{b}k_{c}+\frac{1}{2}%
\epsilon ^{abc}e_{b}e_{c}-2\sigma \epsilon ^{abc}\left( 2\omega _{bd}k_{c}^{%
\text{ }d}+e_{bd}e_{c}^{\text{ }d}\right) =0\,,
\end{eqnarray}%
while the corresponding field equations for the spin-3 fields are%
\begin{eqnarray}
\mathcal{T}^{ab} &\equiv&de^{ab}+\epsilon ^{cd\left( a\right\vert }\omega
_{c}e_{d}^{\text{ }\left\vert b\right) }+\epsilon ^{cd\left( a\right\vert
}e_{c}\omega _{d}^{\text{ }\left\vert b\right) }=0\,, \\
\mathcal{R}^{ab} &\equiv&d\omega ^{ab}+\epsilon ^{cd\left( a\right\vert }\omega
_{c}\omega _{d}^{\text{ }\left\vert b\right) }=0\,, \\
\mathcal{F}^{ab} &\equiv&dk^{ab}+\epsilon ^{cd\left( a\right\vert }\omega
_{c}k_{d}^{\text{ }\left\vert b\right) }+\epsilon ^{cd\left( a\right\vert
}k_{c}\omega _{d}^{\text{ }\left\vert b\right) }+\epsilon ^{cd\left(
a\right\vert }e_{c}e_{d}^{\text{ }\left\vert b\right) }=0\,.\label{hsm3}
\end{eqnarray}
Let us note that the terms $\epsilon ^{abc}e_{b}e_{c}$  and $\epsilon ^{abc}e_{bd}e_{c}^{\text{ }d}$ do not contribute anymore to
the Lorentz curvature equation in the very same way as it happens in the case of the HS flat gravity. Nevertheless, these terms
appear explicitly in the new curvature equation $\mathcal{F}^{a}=0$. \
Analogously, the term $\epsilon ^{cd\left( a\right\vert }e_{c}e_{d}^{\text{ }%
\left\vert b\right) }$ is now in the spin-3 curvature equation $\mathcal{F}%
^{ab}=0$. Thus, the present theory does not only generalize the HS extension of the Poincar\'{e} gravity introducing extra spin-2 and spin-3 fields, but also
modifies the dynamics.
The CS action (\ref{Maxspin3}) is invariant under gauge
transformations $\delta A=D\lambda$, where the gauge parameter is given by
\begin{equation}
\lambda =\Lambda ^{a}M_{a}+\xi ^{a}P_{a}+\chi ^{a}Z_{a}+\Lambda
^{ab}M_{ab}+\xi ^{ab}P_{ab}\,+\chi ^{ab}Z_{ab}\,.
\end{equation}%
For the spin-2 fields we get
\begin{eqnarray}
\delta \omega ^{a} &=&D_{\omega }\Lambda ^{a}\,-4\sigma \epsilon
^{abc}\omega _{bd}\Lambda _{c}^{\text{ }d}\,, \\
\delta e^{a} &=&D_{\omega }\xi ^{a}-\epsilon ^{abc}\Lambda _{b}e_{c}-4\sigma
\epsilon ^{abc}\omega _{bd}\xi _{c}^{\text{ }d}-4\sigma \epsilon
^{abc}e_{bd}\Lambda _{c}^{\text{ }d}\,\,, \\
\delta k^{a} &=&D_{\omega }\chi ^{a}-\epsilon ^{abc}\xi _{b}e_{c}-\epsilon
^{abc}\Lambda _{b}k_{c}-4\sigma \epsilon ^{abc}e_{bd}\xi _{c}^{\text{ }d}
-4\sigma \epsilon ^{abc}\omega _{bd}\chi _{c}^{\text{ }d}-4\sigma \epsilon
^{abc}k_{bd}\Lambda _{c}^{\text{ }d}\,,
\end{eqnarray}%
where, besides the usual gauge transformations of CS Maxwell gravity, there are new terms
proportional to the spin-3 gauge parameters $\xi ^{ab}$, $\Lambda ^{ab}$ and
$\chi ^{ab}$. The spin-3 fields, on the other hand, transform as
\begin{eqnarray}
\delta \omega ^{ab} &=&d\Lambda ^{ab}\,+\epsilon ^{cd\left( a\right\vert
}\omega _{c}\Lambda _{d}^{\text{ }\left\vert b\right) }+\epsilon
^{cd(a}\omega _{\text{ }c}^{b)}\Lambda _{d}\,, \\
\delta e^{ab} &=&d\xi ^{ab}+\epsilon ^{cd\left( a\right\vert }\omega _{c}\xi
_{d}^{\text{ }\left\vert b\right) }+\epsilon ^{cd\left( a\right\vert
}e_{c}\Lambda _{d}^{\text{ }\left\vert b\right) }+\epsilon ^{cd(a}e_{\text{ }%
c}^{b)}\Lambda _{d}+\epsilon ^{cd(a}\omega _{\text{ }c}^{b)}\xi _{d}\,, \\
\delta k^{ab} &=&d\chi ^{ab}+\epsilon ^{cd\left( a\right\vert }k_{c}\Lambda
_{d}^{\text{ }\left\vert b\right) }+\epsilon ^{cd\left( a\right\vert }\omega
_{c}\chi _{d}^{\text{ }\left\vert b\right) }+\epsilon ^{cd\left(
a\right\vert }e_{c}\xi _{d}^{\text{ }\left\vert b\right) }  \notag \\
&&+\,\epsilon ^{cd(a}\omega _{\text{ }c}^{b)}\chi _{d}+\epsilon ^{cd(a}k_{%
\text{ }c}^{b)}\Lambda _{d}+\epsilon ^{cd(a}e_{\text{ }c}^{b)}\xi _{d}\,.
\end{eqnarray}
It is worth to note that the field equation for the field $k^{ab}$ resembles
the structure of the HS matter introduced in \cite{FN, FNN}.

\subsection{$AdS$-Lorentz gravity coupled to spin-3 fields}

Let us consider now the coupling of spin-3 gauge fields to three-dimensional
$AdS$-Lorentz gravity. The abelian semigroup to be considered  in this case is $S_{\mathcal{M}}^{\left( 2\right) }=\left\{ \lambda _{0},\lambda
_{1},\lambda _{2}\right\} $, whose elements satisfy
\begin{equation}
\begin{tabular}{l|lll}
$\lambda _{2}$ & $\lambda _{2}$ & $\lambda _{1}$ & $\lambda _{2}$ \\
$\lambda _{1}$ & $\lambda _{1}$ & $\lambda _{2}$ & $\lambda _{1}$ \\
$\lambda _{0}$ & $\lambda _{0}$ & $\lambda _{1}$ & $\lambda _{2}$ \\ \hline
& $\lambda _{0}$ & $\lambda _{1}$ & $\lambda _{2}$%
\end{tabular}
\label{sm2}
\end{equation}%
Note that the $S_{\mathcal{M}}^{\left( 2\right) }$ semigroup does not have a
zero element as in the $S_{E}^{\left( 2\right) }$ semigroup used in the previous case. The new generators are expressed in term of the original ones
as
\begin{equation*}
\begin{tabular}{lll}
$M_{a}=\lambda _{0}J_{a}\,\,,$ & $\ell P_{a}=\lambda _{1}J_{a}\,,$ & $\text{%
\ }\ell ^{2}Z_{a}=\lambda _{2}J_{a}\,,$ \\
$M_{ab}=\lambda _{0}T_{ab}\,,$ & $\ell P_{ab}=\lambda _{1}T_{ab}\,,$ & $\ell
^{2}Z_{ab}=\lambda _{2}T_{ab}\,.$%
\end{tabular}%
\end{equation*}%
Using the commutation relations of the $\mathfrak{sl}(3,\mathbb{R})$ algebra and the
semigroup multiplication law (\ref{sm2}) one can show that the generators of
the expanded algebra satisfy
\begin{eqnarray}
\left[ M_{a},M_{b}\right]  &=&\epsilon _{abc}M^{c}\,,\text{ \ \ \ \ }\left[
M_{a},P_{b}\right] \;\;=\;\;\epsilon _{abc}P^{c}\,,\text{ \ \ \ \ }\left[ P_{a},P_{b}%
\right] \;\;=\;\;\epsilon _{abc}Z^{c}\,,\, \\
\left[ M_{a},Z_{b}\right]  &=&\epsilon _{abc}Z^{c}\;\;\,\,,\text{ \ \ \ \ }\left[
P_{a},Z_{b}\right] \;\;=\;\;\frac{1}{\ell ^{2}}\epsilon _{abc}P^{c}\,,\text{ \
}\left[ Z_{a},Z_{b}\right] \;\;=\;\;\frac{1}{\ell ^{2}}\epsilon _{abc}Z^{c}\,,
\end{eqnarray}%
\begin{eqnarray}
\left[ M_{a},M_{bc}\right]  &=&\epsilon _{\text{ \ }a\left( b\right.
}^{m}M_{\left. c\right) m}\,,\text{ \ \ \ \ }\left[ M_{a},P_{bc}\right]
\;\;=\;\;\epsilon _{\text{ \ }a\left( b\right. }^{m}P_{\left. c\right) m}\,, \\
\left[ P_{a},M_{bc}\right]  &=&\epsilon _{\text{ \ }a\left( b\right.
}^{m}P_{\left. c\right) m}\,,\text{ \ \ \ \ \ \ }\left[ P_{a},P_{bc}\right]
\;\;=\;\;\epsilon _{\text{ \ }a\left( b\right. }^{m}Z_{\left. c\right) m}\,, \\
\left[ Z_{a},M_{bc}\right]  &=&\epsilon _{\text{ \ }a\left( b\right.
}^{m}Z_{\left. c\right) m}\,,\text{ \ \ \ \ \ \ }\left[ Z_{a},P_{bc}\right] \;\;=\;\;%
\frac{1}{\ell ^{2}}\epsilon _{\text{ \ }a\left( b\right. }^{m}P_{\left.
c\right) m}\,, \\
\left[ M_{a},Z_{bc}\right]  &=&\epsilon _{\text{ \ }a\left( b\right.
}^{m}Z_{\left. c\right) m}\,\,,\text{ \ \ \ \ \ }\left[ P_{a},Z_{bc}\right] \;\;=\;\;%
\frac{1}{\ell ^{2}}\epsilon _{\text{ \ }a\left( b\right. }^{m}P_{\left.
c\right) m}\,, \\
\left[ Z_{a},Z_{bc}\right]  &=&\frac{1}{\ell ^{2}}\epsilon _{\text{ \ }%
a\left( b\right. }^{m}Z_{\left. c\right) m}\,,
\end{eqnarray}%
\begin{eqnarray}
\left[ M_{ab},M_{cd}\right]  &=&\sigma \left( \eta _{a\left( c\right.
}\epsilon _{\left. d\right) bm}+\eta _{b\left( c\right. }\epsilon _{\left.
d\right) am}\right) M^{m}\,, \\
\left[ M_{ab},P_{cd}\right]  &=&\sigma \left( \eta _{a\left( c\right.
}\epsilon _{\left. d\right) bm}+\eta _{b\left( c\right. }\epsilon _{\left.
d\right) am}\right) P^{m}\,, \\
\left[ M_{ab},Z_{cd}\right]  &=&\sigma \left( \eta _{a\left( c\right.
}\epsilon _{\left. d\right) bm}+\eta _{b\left( c\right. }\epsilon _{\left.
d\right) am}\right) Z^{m}\,, \\
\left[ P_{ab},P_{cd}\right]  &=&\sigma \left( \eta _{a\left( c\right.
}\epsilon _{\left. d\right) bm}+\eta _{b\left( c\right. }\epsilon _{\left.
d\right) am}\right) Z^{m}\,, \\
\left[ P_{ab},Z_{cd}\right]  &=&\frac{\sigma }{\ell ^{2}}\left( \eta
_{a\left( c\right. }\epsilon _{\left. d\right) mb}+\eta _{b\left( c\right.
}\epsilon _{\left. d\right) ma}\right) P^{m}\,, \\
\left[ Z_{ab},Z_{cd}\right]  &=&\frac{\sigma }{\ell ^{2}}\left( \eta
_{a\left( c\right. }\epsilon _{\left. d\right) mb}+\eta _{b\left( c\right.
}\epsilon _{\left. d\right) ma}\right) Z^{m}\,.
\end{eqnarray}%
This algebra describes the coupling of the spin-3 generators $\left\{
M_{ab},P_{ab},Z_{ab}\right\} $ to the $AdS$-Lorentz algebra \cite%
{Sorokas, SS} (also known as $\mathfrak{C}_{4}$ \cite{CDMR}) spanned by $\left\{
M_{a},P_{a},Z_{a}\right\} $. It can be written as
three copies of the $\mathfrak{sl}\left( 3,%
\mathbb{R}
\right) $ when the generators are redefined as
\begin{eqnarray}
J_{a} &=&\frac{1}{2}\left( \ell ^{2}Z_{a}+\ell P_{a}\right) \,,\text{ \ \ \
\ \ \ }\bar{J}_{a}\;\;=\;\;\frac{1}{2}\left( \ell ^{2}Z_{a}-\ell P_{a}\right) \,,%
\text{ \ \ \ \ \ \ }\tilde{J}_{a}\;\;=\;\;M_{a}-\ell ^{2}Z_{a}\,, \\
T_{ab} &=&\frac{1}{2}\left( \ell ^{2}Z_{ab}+\ell P_{ab}\right) \,,\text{ \ \ \
}\bar{T}_{ab}\;\;=\;\;\frac{1}{2}\left( \ell ^{2}Z_{ab}-\ell P_{ab}\right) \,,%
\text{ \ \ \ }\tilde{T}_{ab}\;\;=\;\;M_{ab}-\ell ^{2}Z_{ab}\,.
\end{eqnarray}
This new spin-3 symmetry is quite different from the Maxwell one although
having the same number of spin-2 and spin-3 generators. In particular, there
are no abelian generators and the parameter $\ell $ appears explicitly in
some commutation relations. Interestingly, the spin-3 Maxwell algebra is
recovered in the limit $\ell \rightarrow \infty $. Something similar happens with the quadratic Casimir associated to this algebra, which can be deduced using the expansion method and reads
\begin{eqnarray}
C &=&\mu _{0}\left( M_{a}M^{a}-\frac{1}{2\sigma }M_{ab}M^{ab}\right) +2\frac{%
\mu _{1}}{\ell }\left[ P_{a}M^{a}+\frac{1}{\ell }Z_{a}P^{a}-\frac{1}{2\sigma
}\left( P_{ab}M^{ab}+\frac{1}{\ell }P_{ab}Z^{ab}\right) \right]   \notag \\
&&+\frac{\mu _{2}}{\ell ^{2}}\left[ P_{a}P^{a}+2Z_{a}M^{a}+\frac{1}{\ell ^{2}%
}Z_{a}Z^{a}-\frac{1}{2\sigma }\left( P_{ab}P^{ab}+2Z_{ab}M^{ab}+\frac{1}{%
\ell ^{2}}Z_{ab}Z^{ab}\right) \right] \,,
\end{eqnarray}%
In this case, however, in order to define a flat limit it is necessary to redefine the constants $\mu_1$ and $\mu_2$ in the form
\begin{equation}
\mu _{1}\rightarrow \ell \mu _{1}\,,\text{ \ \ }\mu _{2}\rightarrow \ell
^{2}\mu _{2}\,.  \label{rescadsl}
\end{equation}%
Then, the Casimir operator of the spin-3 extension of the Maxwell algebra (\ref{invmax}) is recovered in the limit $%
\ell \rightarrow \infty $.

Let us consider now the one-form gauge connection%
\begin{equation}
A=\omega ^{a}M_{a}+e^{a}P_{a}+k^{a}Z_{a}+\omega
^{ab}M_{ab}+e^{ab}P_{ab}+k^{ab}Z_{ab}\text{ ,}  \label{adsloc}
\end{equation}%
and the invariant tensor that follows from the expansion
\begin{eqnarray}
\left\langle M_{a}M_{b}\right\rangle  &=&\frac{\mu _{0}}{2}\eta _{ab}\,,%
\text{ \ \ }\left\langle M_{a}P_{b}\right\rangle \;\;=\;\;\frac{\mu _{1}}{2\ell }%
\eta _{ab}\,,\,\;\;\ \ \left\langle P_{a}P_{b}\right\rangle \;\;=\;\;\frac{\mu _{2}}{%
2\ell ^{2}}\eta _{ab}\,, \\
\left\langle M_{a}Z_{b}\right\rangle  &=&\frac{\mu _{2}}{2\ell ^{2}}\eta
_{ab}\,,\text{ \ \ }\left\langle P_{a}Z_{b}\right\rangle \;\;=\;\;\frac{\mu _{1}}{%
2\ell ^{3}}\eta _{ab}\,,\text{ \ \ }\left\langle Z_{a}Z_{b}\right\rangle \;\;=\;\;%
\frac{\mu _{2}}{2\ell ^{4}}\eta _{ab}\,,
\end{eqnarray}%
\begin{eqnarray}
\left\langle M_{ab}M_{bc}\right\rangle  &=&-\mu _{0}\frac{\sigma }{2}\left(
\eta _{a(c}\eta _{d)b}-\frac{2}{3}\eta _{ab}\eta _{dc}\right) \,, \\
\left\langle P_{ab}P_{bc}\right\rangle  &=&-\frac{\mu _{2}}{\ell ^{2}}\frac{%
\sigma }{2}\left( \eta _{a(c}\eta _{d)b}-\frac{2}{3}\eta _{ab}\eta
_{dc}\right) \,, \\
\left\langle M_{ab}P_{bc}\right\rangle  &=&-\frac{\mu _{1}}{\ell }\frac{%
\sigma }{2}\left( \eta _{a(c}\eta _{d)b}-\frac{2}{3}\eta _{ab}\eta
_{dc}\right) \,, \\
\left\langle M_{ab}Z_{bc}\right\rangle  &=&-\frac{\mu _{2}}{\ell ^{2}}\frac{%
\sigma }{2}\left( \eta _{a(c}\eta _{d)b}-\frac{2}{3}\eta _{ab}\eta
_{dc}\right) \,, \\
\left\langle P_{ab}Z_{bc}\right\rangle  &=&-\frac{\mu _{1}}{\ell ^{3}}\frac{%
\sigma }{2}\left( \eta _{a(c}\eta _{d)b}-\frac{2}{3}\eta _{ab}\eta
_{dc}\right) \,, \\
\left\langle Z_{ab}Z_{bc}\right\rangle  &=&-\frac{\mu _{2}}{\ell ^{4}}\frac{%
\sigma }{2}\left( \eta _{a(c}\eta _{d)b}-\frac{2}{3}\eta _{ab}\eta
_{dc}\right) \,.
\end{eqnarray}%
The CS action (\ref{eq:action}) in this case takes the form
\begin{eqnarray}
S &=&\kappa \int \mu _{0}\left[ \frac{1}{2}\left( \omega ^{a}d\omega
_{a}+\frac{1}{3}\epsilon _{abc}\omega ^{a}\omega ^{b}\omega ^{c}\right)
-\sigma \left( \omega _{\text{ }b}^{a}d\omega _{\text{ }a}^{b}+2\epsilon
_{abc}\omega ^{a}\omega ^{bd}\omega _{\text{ }d}^{c}\right) \right]   \notag
\\
&&+\frac{\mu _{1}}{\ell }\left[ e^{a}\left( d\omega _{a}+\frac{1}{2}\epsilon
_{abc}\left\{ \omega ^{b}\omega ^{c}+\frac{1}{2\ell ^{4}}k^{b}k^{c}\right\} +%
\frac{1}{\ell ^{2}}dk_{a}+\frac{1}{\ell ^{2}}\epsilon _{abc}\left\{ \omega
^{b}k^{c}+\frac{1}{\ell ^{4}}k^{b}k^{c}\right\} \right) \right.   \notag \\
&&\left. +\frac{1}{6\ell ^{2}}\epsilon _{abc}\left( e^{a}e^{b}e^{c}-12\sigma
\epsilon _{abc}e^{a}e^{bd}e_{\text{ }d}^{c}\right) -2\sigma e^{a}\left(
\epsilon _{abc}\omega ^{bd}\omega _{\text{ }d}^{c}+\frac{2}{\ell ^{2}}%
\epsilon _{abc}\omega ^{bd}k_{\text{ }d}^{c}\right) \right.   \notag \\
&&\left. -2\sigma e^{ab}\left( d\omega _{ab}+2\epsilon _{acd}\left\{ \omega
^{c}\omega _{\ b}^{d}+\frac{1}{\ell ^{2}}k^{c}\omega _{\ b}^{d}+\frac{1}{%
\ell ^{4}}k^{c}k_{\ b}^{d}\right\} +\frac{1}{\ell ^{2}}dk_{ab}\right.
\right.   \notag \\
&&\left. \left. +\frac{2}{\ell ^{2}}\epsilon _{acd}\left\{ k^{c}\omega _{\
b}^{d}+2\omega ^{c}k_{\ b}^{d}+\frac{2}{\ell ^{2}}k^{c}k_{\ b}^{d}\right\}
\right) \right]   \notag \\
&&+\frac{\mu _{2}}{\ell ^{2}}\left[ \frac{1}{2}e^{a}\left( de_{a}+\epsilon
_{abc}\omega ^{b}e^{c}+\frac{1}{\ell ^{2}}\epsilon _{abc}k^{b}e^{c}\right)
\right.   \notag \\
&&\left. +k^{a}\left( d\omega _{a}+\frac{1}{2}\epsilon _{abc}\omega
^{b}\omega ^{c}+\frac{1}{\ell ^{2}}\left\{ dk_{a}+\frac{1}{2}\epsilon
_{abc}\omega ^{b}k^{c}+\frac{1}{3\ell ^{2}}\epsilon _{abc}k^{b}k^{c}\right\}
\right) \right.   \notag \\
&&\left. -2\sigma \left( \omega ^{ab}dk_{ab}+\frac{1}{2\ell ^{2}}%
k^{ab}dk_{ab}+\epsilon _{abc}k^{a}\omega ^{be}\omega _{\text{ }e}^{c}+\frac{2%
}{\ell ^{2}}\epsilon _{abc}k^{a}\omega ^{be}k_{\text{ }e}^{c}\right. \right.
\notag \\
&&\left. \left. +\frac{1}{\ell ^{4}}\epsilon _{abc}k^{a}k^{be}k_{\text{ }%
e}^{c}+2\epsilon _{abc}\omega ^{a}k^{be}\omega _{\text{ }e}^{c}+\frac{1}{%
\ell ^{2}}\epsilon _{abc}\omega ^{a}k^{be}k_{\text{ }e}^{c}\right) \right.
\notag \\
&&\left. -\sigma e_{\text{ }}^{ab}\left( de_{ab}+2\epsilon _{acd}\omega
^{c}e_{\text{ }b}^{d}+\frac{2}{\ell ^{2}}\epsilon _{acd}k^{c}e_{\text{ }%
b}^{d}+4\epsilon _{acd}e^{c}\omega _{\text{ }b}^{d}+\frac{4}{\ell ^{2}}%
\epsilon _{acd}e^{c}k_{\text{ }b}^{d}\right) \right] \,.  \label{AdSlspin3}
\end{eqnarray}%
and describes the coupling of spin-3 fields to the $AdS$-Lorentz
gravity.
Note that the absence of abelian generators in this new HS symmetry gives terms proportional to $\mu _{1}$ and $\mu _{2}~$ that are different to the ones that appear in the spin-3 Maxwell case. However, the CS action (\ref{Maxspin3}) describing
the coupling of spin-3 fields to the Maxwell gravity can be recovered
considering the redefinition (\ref{rescadsl}) of the constants and applying
the flat limit $\ell \rightarrow \infty $. The same procedure can be done at
the level of the invariant tensor to obtain the one corresponding to the spin-3 Maxwell algebra.

The field equations in this case are given by%
\begin{eqnarray}
\mathcal{T}^{a} &\equiv &de^{a}+\epsilon ^{abc}\left( \omega _{b}e_{c}+\frac{%
1}{\ell ^{2}}k_{b}e_{c}\right) -4\sigma \epsilon ^{abc}\left( e^{bd}\omega
_{c}^{\text{ }d}+\frac{1}{\ell ^{2}}e^{bd}k_{c}^{\text{ }d}\right) =0\,, \\
\mathcal{R}^{a} &\equiv &d\omega ^{a}+\frac{1}{2}\epsilon ^{abc}\omega
_{b}\omega _{c}-2\sigma \epsilon ^{abc}\omega _{bd}\omega _{c}^{\text{ }%
d}=0\,, \\
\mathcal{F}^{a} &\equiv &dk^{a}+\epsilon ^{abc}\left( \omega _{b}k_{c}+\frac{%
1}{\ell ^{2}}k_{b}k_{c}+\frac{1}{2}e_{b}e_{c}\right) \,-2\sigma \epsilon
^{abc}\left( 2\omega _{bd}k_{c}^{\text{ }d}+\frac{1}{\ell ^{2}}k_{bd}k_{c}^{%
\text{ }d}+e_{bd}e_{c}^{\text{ }d}\right) =0\,,\\
\mathcal{T}^{ab} &\equiv&de^{ab}+\epsilon ^{cd\left( a\right\vert }\left( \omega
_{c}e_{d}^{\text{ }\left\vert b\right) }+\frac{1}{\ell ^{2}}k_{c}e_{d}^{%
\text{ }\left\vert b\right) }+e_{c}\omega _{d}^{\text{ }\left\vert b\right)
}+\frac{1}{\ell ^{2}}e_{c}k_{d}^{\text{ }\left\vert b\right) }\right) =0\,,
\\
\mathcal{R}^{ab} &\equiv&d\omega ^{ab}+\epsilon ^{cd\left( a\right\vert }\omega
_{c}\omega _{d}^{\text{ }\left\vert b\right) }=0\,, \\
\mathcal{F}^{ab} &\equiv&dk^{ab}+\epsilon ^{cd\left( a\right\vert }\left( \omega
_{c}k_{d}^{\text{ }\left\vert b\right) }+\frac{1}{\ell ^{2}}k_{c}k_{d}^{%
\text{ }\left\vert b\right) }+k_{c}\omega _{d}^{\text{ }\left\vert b\right)
}+\frac{1}{\ell ^{2}}k_{c}k_{d}^{\text{ }\left\vert b\right) }+e_{c}e_{d}^{%
\text{ }\left\vert b\right) }\right) =0\,,
\end{eqnarray}%
where presence of non-abelian generators also modifies the gauge
transformations w.r.t the spin-3 Maxwell case. Indeed, the spin-2 gauge
transformations have the form
\begin{eqnarray}
\delta \omega ^{a} &=&D_{\omega }\Lambda ^{a}\,-4\sigma \epsilon
^{abc}\omega _{bd}\Lambda _{c}^{\text{ }d}\,, \\
\delta e^{a} &=&D_{\omega }\xi ^{a}+\frac{1}{\ell ^{2}}\epsilon
^{abc}k_{b}\xi _{c}-\epsilon ^{abc}\Lambda _{b}e_{c}-\frac{1}{\ell ^{2}}%
\epsilon ^{abc}\chi _{b}e_{c}  \notag \\
&&-4\sigma \epsilon ^{abc}\left( \omega _{bd}\xi _{c}^{\text{ }d}+\frac{1}{%
\ell ^{2}}k_{bd}\xi _{c}^{\text{ }d}+e_{bd}\Lambda _{c}^{\text{ }d}\,+\frac{1%
}{\ell ^{2}}e_{bd}\chi _{c}^{\text{ }d}\right) , \\
\delta k^{a} &=&D_{\omega }\chi ^{a}+\frac{1}{\ell ^{2}}\epsilon
^{abc}k_{b}\chi _{c}-\epsilon ^{abc}\xi _{b}e_{c}-4\sigma \epsilon
^{abc}\left( e_{bd}\xi _{c}^{\text{ }d}+k_{bd}\chi _{c}^{\text{ }d}\right)
\,,
\end{eqnarray}%
while the spin-3 gauge transformations are given by%
\begin{eqnarray}
\delta \omega ^{ab} &=&d\Lambda ^{ab}\,+\epsilon ^{cd\left( a\right\vert
}\omega _{c}\Lambda _{d}^{\text{ }\left\vert b\right) }+\epsilon
^{cd(a}\omega _{\text{ }c}^{b)}\Lambda _{d}\,, \\
\delta e^{ab} &=&d\xi ^{ab}+\epsilon ^{cd\left( a\right\vert }\omega _{c}\xi
_{d}^{\text{ }\left\vert b\right) }+\epsilon ^{cd(a}\omega _{\text{ }%
c}^{b)}\xi _{d}+\frac{1}{\ell ^{2}}\epsilon ^{cd\left( a\right\vert
}k_{c}\xi _{d}^{\text{ }\left\vert b\right) }+\frac{1}{\ell ^{2}}\epsilon
^{cd(a}k_{\text{ }c}^{b)}\xi _{d}  \notag \\
&&+\epsilon ^{cd\left( a\right\vert }e_{c}\Lambda _{d}^{\text{ }\left\vert
b\right) }+\frac{1}{\ell ^{2}}\epsilon ^{cd\left( a\right\vert }e_{c}\chi
_{d}^{\text{ }\left\vert b\right) }+\epsilon ^{cd(a}e_{\text{ }%
c}^{b)}\Lambda _{d}+\frac{1}{\ell ^{2}}\epsilon ^{cd(a}e_{\text{ }%
c}^{b)}\chi _{d}\,, \\
\delta k^{ab} &=&d\chi ^{ab}\,+\epsilon ^{cd\left( a\right\vert }\omega
_{c}\chi _{d}^{\text{ }\left\vert b\right) }+\epsilon ^{cd(a}\omega _{\text{
}c}^{b)}\chi _{d}+\frac{1}{\ell ^{2}}\epsilon ^{cd\left( a\right\vert
}k_{c}\chi _{d}^{\text{ }\left\vert b\right) }+\frac{1}{\ell ^{2}}\epsilon
^{cd(a}k_{\text{ }c}^{b)}\chi _{d}  \notag \\
&&+\epsilon ^{cd\left( a\right\vert }e_{c}\xi _{d}^{\text{ }\left\vert
b\right) }+\epsilon ^{cd(a}e_{\text{ }c}^{b)}\xi _{d}\,.
\end{eqnarray}%
Note that the limit $\ell \rightarrow \infty $ properly reproduces the
spin-3 Maxwell field equations and gauge transformations.

\section{Generalizations to $\mathfrak{B}_{m}$ and $\mathfrak{C}_{m}$
gravity theories}

A wide class of expanded Lie algebras have recently been introduced in the context of CS gravity in diverse dimensions. In particular, the $\mathfrak{B}_{m}$ family has been
useful in order to recover standard General Relativity without cosmological
constant as a particular limit of a CS and Born-Infeld like
gravity theories \cite{EHTZ, GRCS, CPRS1, CPRS2, CPRS3}. Subsequently, the $%
\mathfrak{C}_{m}$ algebras were used to recover diverse Pure Lovelock
gravity actions in a matter-free configuration limit \cite{CDIMR, CMR, CR}.
Furthermore, it was shown in \cite{CDMR, Durka}, that the $AdS$ and Poincar\'{e} algebra, correspond to the simplest cases ($m=3$) of the $\mathfrak{C}_{m}$ family and the $\mathfrak{B}_{m}$ family, respectively. Along the same line, the $AdS$-Lorentz and its In\"{o}n\"{u}-Wigner contraction, the Maxwell algebra, are given by the $\mathfrak{C}_{4}$ and the $\mathfrak{B}%
_{4}$ algebras, respectively. Generically, the $\mathfrak{B}_{m}$
symmetries can always be recovered as an In\"{o}n\"{u}-Wigner contraction of the $\mathfrak{C}%
_{m}$ symmetries.

In this section, we extend this construction to present new non-trivial Lie algebras that correspond to the spin-3 extensions of the $\mathfrak{B}_{m}$ and $\mathfrak{C}_{m}$ algebras. In every case, the new
commutation relations can be obtained as an expansion of the $\mathfrak{%
sl}\left( 3,%
\mathbb{R}
\right) $ algebra considering two families of semigroups. In addition,
we study the three-dimensional CS action based on these new HS symmetries.
Finally, we provide with the general limit relating such new symmetries.

\subsection{$\mathfrak{B}_{m}$ gravity coupled to spin-3 fields}
Let us consider the semigroup $S_{E}^{\left( m-2\right) }=\left\{ \lambda _{0},\lambda
_{1},\dots ,\lambda _{m-1}\right\} $, whose multiplication law is given by
\begin{equation}
\lambda _{\alpha }\lambda _{\beta }=\left\{
\begin{array}{c}
\lambda _{\alpha +\beta },\text{ \ \ \ \ if }\alpha +\beta \leq m-1 \\
\lambda _{m-1},\text{ \ \ \ \ if }\alpha +\beta >m-1%
\end{array}
\right. ,
\end{equation}
where $\lambda _{m-1}=0_{S}$ is the zero element of the semigroup. The $S_{E}^{\left(
m-2\right) }$-expanded algebra is spanned by the set of generators%
\begin{equation}
\mathfrak{B}_{m}^{s_{3}}=S_{E}^{\left( m-2\right) }\times \mathfrak{sl}\left( 3,\mathbb{R}\right)%
=\left\{ M_{a}^{\left( i\right) },P_{a}^{\left( \bar{\imath}\right)
},M_{ab}^{\left( i\right) },P_{ab}^{\left( \bar{\imath}\right) }\right\} \,.,
\end{equation}%
which are related to the original ones by%
\begin{eqnarray}
\ell ^{i}M_{a}^{\left( i\right) } &=&J_{\left( a,i\right) }\;\;=\;\;\lambda
_{i}J_{a}\,,\text{ \ \ }\ell ^{i}M_{ab}^{\left( i\right) }\;\;=\;\;T_{\left(
ab,i\right) }\;\;=\;\;\lambda _{i}T_{ab}\,, \\
\ell ^{\bar{\imath}}P_{a}^{\left( \bar{\imath}\right) } &=&J_{\left( a,\bar{%
\imath}\right) }\;\;=\;\;\lambda _{\bar{\imath}}J_{a}\,,\text{ \ \ \ }\ell ^{\bar{%
\imath}}P_{ab}^{\left( \bar{\imath}\right) }\;\;=\;\;T_{\left( ab,\bar{\imath}%
\right) }\;\;=\;\;\lambda _{\bar{\imath}}T_{ab}\,.
\end{eqnarray}%
Here $i$ takes even values and $\bar{\imath}$ takes odd values. The spin-3
and spin-2 generators of the expanded algebra satisfy%
\begin{eqnarray}
\left[ M_{a}^{\left( i\right) },M_{b}^{\left( j\right) }\right] &=&\epsilon
_{abc}M^{c,\left( i+j\right) },\,\,\,\,\mathrm{for}\,\,\,\,i+j\leq m-2\,, \\
\left[ M_{a}^{\left( i\right) },P_{b}^{\left( \bar{\imath}\right) }\right]
&=&\epsilon _{abc}P^{c,\left( i+\bar{\imath}\right) },\,\ \,\,\,\mathrm{for}%
\,\,\,\,i+\bar{\imath}\leq m-2\,, \\
\left[ P_{a}^{\left( \bar{\imath}\right) },P_{b}^{\left( \bar{j}\right) }%
\right] &=&\epsilon _{abc}M^{c,\left( \bar{\imath}+\bar{j}\right) },\,\,\,\,%
\mathrm{for}\,\,\,\,\bar{\imath}+\bar{j}\leq m-2\,,
\end{eqnarray}%
\begin{eqnarray}
\left[ M_{a}^{\left( i\right) },M_{bc}^{\left( j\right) }\right] &=&\epsilon
_{\text{ \ }a\left( b\right. }^{m}M_{\left. c\right) m}^{\left( i+j\right)
}\,,\,\,\,\,\mathrm{for}\,\,\,\,i+j\leq m-2\,, \\
\left[ M_{a}^{\left( i\right) },P_{bc}^{\left( \bar{\imath}\right) }\right]
&=&\epsilon _{\text{ \ }a\left( b\right. }^{m}P_{\left. c\right) m}^{\left(
i+\bar{\imath}\right) }\,,\,\,\ \,\,\mathrm{for}\,\,\,\,i+\bar{\imath}\leq
m-2\,, \\
\left[ P_{a}^{\left( \bar{\imath}\right) },M_{bc}^{\left( j\right) }\right]
&=&\epsilon _{\text{ \ }a\left( b\right. }^{m}P_{\left. c\right) m}^{\left(
\bar{\imath}+j\right) }\,,\,\,\,\ \,\mathrm{for}\,\,\,\,\bar{\imath}+j\leq
m-2\,, \\
\left[ P_{a}^{\left( \bar{\imath}\right) },P_{bc}^{\left( \bar{j}\right) }%
\right] &=&\epsilon _{\text{ \ }a\left( b\right. }^{m}M_{\left. c\right)
m}^{\left( \bar{\imath}+\bar{j}\right) }\,,\,\,\,\,\mathrm{for}\,\,\,\,\bar{%
\imath}+\bar{j}\leq m-2\,,
\end{eqnarray}%
\begin{eqnarray}
\left[ M_{ab}^{\left( i\right) },M_{cd}^{\left( j\right) }\right] &=&\sigma
\left( \eta _{a\left( c\right. }\epsilon _{\left. d\right) bm}+\eta
_{b\left( c\right. }\epsilon _{\left. d\right) am}\right) M^{m,\left(
i+j\right) },\,\,\,\,\mathrm{for}\,\,\,\,i+j\leq m-2\,, \\
\left[ M_{ab}^{\left( i\right) },P_{cd}^{\left( \bar{\imath}\right) }\right]
&=&\sigma \left( \eta _{a\left( c\right. }\epsilon _{\left. d\right)
bm}+\eta _{b\left( c\right. }\epsilon _{\left. d\right) am}\right)
P^{m,\left( i+\bar{\imath}\right) },\,\ \,\,\,\mathrm{for}\,\,\,\,i+\bar{%
\imath}\leq m-2\,, \\
\left[ P_{ab}^{\left( i\right) },P_{cd}^{\left( \bar{j}\right) }\right]
&=&\sigma \left( \eta _{a\left( c\right. }\epsilon _{\left. d\right)
bm}+\eta _{b\left( c\right. }\epsilon _{\left. d\right) am}\right)
M^{m,\left( \bar{\imath}+\bar{j}\right) },\,\,\,\,\mathrm{for}\,\,\,\,\bar{%
\imath}+\bar{j}\leq m-2\,, \\
\text{\textrm{others }} &=&0\,.
\end{eqnarray}%
The new algebra, which we denote by $\mathfrak{B}_{m}^{s_{3}}$,
describes the coupling of the spin-3 generators $\left\{ M_{ab}^{\left( i\right)
},P_{ab}^{\left( \bar{\imath}\right) }\right\} $ to the spin-2 generators $%
\left\{ M_{a}^{\left( i\right) },P_{a}^{\left( \bar{\imath}\right) }\right\}
$ of the algebra $\mathfrak{B}_{m}$. Let us remark that $m=3,\,4$ reproduce
the HS extension of the Poincar\'{e} and Maxwell algebras discussed in the previous sections.
The quadratic Casimir in this case is given by
\begin{eqnarray}
C &=&\mu _{i+j}\left( \sum_{i,j}^{i+j\leq m-2}M_{a}^{\left( i\right)
}M^{a,\left( j\right) }-\frac{1}{2\sigma }M_{ab}^{\left( i\right)
}M^{ab,\left( j\right) }\right) +\mu _{i+\bar{\imath}}\left( \sum_{i,\bar{\imath}}^{i+\bar{\imath}\leq
m-2}M_{a}^{\left( i\right) }P^{a,\left( \bar{\imath}\right) }-\frac{1}{%
2\sigma }M_{ab}^{\left( i\right) }P^{ab,\left( \bar{\imath}\right) }\right)
\notag \\
&&+\mu _{\bar{\imath}+\bar{j}}\left( \sum_{\bar{\imath},\bar{j}}^{\bar{\imath%
}+\bar{j}\leq m-2}\,P_{a}^{\left( \bar{\imath}\right) }P^{a,\left( \bar{j}%
\right) }-\frac{1}{2\sigma }P_{ab}^{\left( \bar{\imath}\right) }P^{ab,\left(
\bar{j}\right) }\right) \,,
\end{eqnarray}%
while the invariant tensor reads
\begin{eqnarray}
\left\langle M_{a}^{\left( i\right) }M_{b}^{\left( j\right) }\right\rangle
&=&\frac{1}{2}\mu _{i+j}\,\eta _{ab}\,,\text{ \ \ }\mathrm{for}%
\,\,\,\,i+j\leq m-2\,, \\
\left\langle M_{a}^{\left( i\right) }P_{b}^{\left( \bar{\imath}\right)
}\right\rangle &=&\frac{1}{2}\mu _{i+\bar{\imath}}\,\eta _{ab}\,,\text{ \ \ }%
\mathrm{for}\,\,\,\,i+\bar{\imath}\leq m-2\,, \\
\left\langle P_{a}^{\left( \bar{\imath}\right) }P_{b}^{\left( \bar{j}\right)
}\right\rangle &=&\frac{1}{2}\mu _{\bar{\imath}+\bar{j}}\,\eta _{ab}\,,\,\ \
\ \mathrm{for}\,\,\,\,\bar{\imath}+\bar{j}\leq m-2\,, \\
\left\langle M_{ab}^{\left( i\right) }M_{cd}^{\left( j\right) }\right\rangle
&=&-\frac{\sigma }{2}\mu _{i+j}\,\left( \eta _{a\left( c\right. }\epsilon
_{\left. d\right) b}-\frac{2}{3}\eta _{ab}\eta _{cd}\right) \,,\,\ \ \
\mathrm{for}\,\,\,\,i+j\leq m-2\,, \\
\left\langle M_{ab}^{\left( i\right) }P_{cd}^{\left( \bar{\imath}\right)
}\right\rangle &=&-\frac{\sigma }{2}\mu _{i+\bar{\imath}}\,\left( \eta
_{a\left( c\right. }\epsilon _{\left. d\right) b}-\frac{2}{3}\eta _{ab}\eta
_{cd}\right) \,,\,\ \ \ \mathrm{for}\,\,\,\,i+\bar{\imath}\leq m-2\,, \\
\left\langle P_{ab}^{\left( \bar{\imath}\right) }P_{cd}^{\left( \bar{j}%
\right) }\right\rangle &=&-\frac{\sigma }{2}\mu _{\bar{\imath}+\bar{j}%
}\,\left( \eta _{a\left( c\right. }\epsilon _{\left. d\right) b}-\frac{2}{3}%
\eta _{ab}\eta _{cd}\right) \,,\,\ \ \ \mathrm{for}\,\,\,\,\bar{\imath}+\bar{%
j}\leq m-2\,.
\end{eqnarray}%

By introducing the $\mathfrak{B}_{m}^{s_{3}}$-valued connection one-form
\begin{equation}
A=\omega ^{a,\left( i\right) }M_{a}^{\left( i\right) }+e^{a,\left( \bar{%
\imath}\right) }P_{a}^{\left( \bar{\imath}\right) }+\omega ^{ab,\left(
i\right) }M_{ab}^{\left( i\right) }+e^{ab,\left( \bar{\imath}\right)
}P_{ab}^{\left( \bar{\imath}\right) }\,,  \label{ofbk}
\end{equation}%
a CS action (\ref{eq:action}) can be constructed:
\begin{eqnarray}
S &=&\kappa \int \mu _{i+j}\,\left[ \frac{1}{2}\left( \omega
^{a,\left( i\right) }d\omega _{a}^{\left( j\right) }+\frac{1}{3}\epsilon
_{abc}\omega ^{a,\left( l\right) }\omega ^{b,\left( m\right) }\omega
^{c,\left( n\right) }\,\delta _{l+m+n}^{i+j}\right) \right.   \notag \\
&&\left. -\sigma \left( \omega _{\text{ }b}^{a,\left( i\right) }d\omega _{%
\text{ }a}^{b,\left( j\right) }+2\epsilon _{abc}\omega ^{a,\left( l\right)
}\omega ^{bd,\left( m\right) }\omega _{\text{ }d}^{c,\left( n\right)
}\,\delta _{l+m+n}^{i+j}\right) \right]   \notag \\
&&+\mu _{i+\bar{\imath}}\left[ e^{a,\left( \bar{\imath}\right) }\left(
d\omega _{a}^{\left( i\right) }+\frac{1}{2}\epsilon _{abc}\omega ^{b,\left(
m\right) }\omega ^{c,\left( n\right) }\,\delta _{m+n}^{i}-2\sigma \epsilon
_{abc}\omega ^{bd,\left( m\right) }\omega _{\text{ }d}^{c,\left( n\right)
}\delta _{m+n}^{i}\right) \right.   \notag \\
&&\left. -2\sigma e^{ab,\left( \bar{\imath}\right) }\left( d\omega _{\text{ }%
a}^{b,\left( i\right) }+2\epsilon _{acd}\omega ^{c,\left( m\right) }\omega _{%
\text{ }b}^{d,\left( n\right) }\,\delta _{m+n}^{i}\right) \right]   \notag \\
&&+\mu _{\bar{\imath}+\bar{j}}\left[ \frac{1}{2}e^{a,\left( \bar{\imath}%
\right) }\left( de_{a}^{\left( \bar{j}\right) }+\epsilon _{abc}\omega
^{b,\left( m\right) }e^{c,\left( \bar{n}\right) }\delta _{m+\bar{n}}^{\bar{j}%
}\right) \right.   \notag \\
&&\left. -\sigma e_{\text{ }}^{ab,\left( \bar{\imath}\right) }\left(
de_{ab}^{\left( \bar{j}\right) }+2\epsilon _{acd}\omega ^{c,\left( m\right)
}e_{\text{ }b}^{d,\left( \bar{n}\right) }\delta _{m+\bar{n}}^{\bar{j}%
}+4\epsilon _{acd}e^{c,\left( \bar{m}\right) }\omega _{\text{ }b}^{d,\left(
n\right) }\delta _{\bar{m}+n}^{\bar{j}}\right) \right] \,,  \label{AccBk}
\end{eqnarray}%
and describes the coupling of spin-3 gauge fields to $\mathfrak{B}%
_{m}$ gravity. It is important to clarify that the constants $\mu _{p+q}$ are
well-defined only for $p+q\leq m-2$. In particular, the Poincar\'{e} gravity
action coupled to spin-3 gauge fields is recovered for $m=3$ (see eq. (\ref%
{poinc3})). New gravity actions coupled to spin-3 fields appear for $m\geq 4$,
where for the $m=4$ we recover the Maxwell case (see eq. (\ref{Maxspin3})). It is important to emphasize that the
procedure considered here allows us to build the most general CS action
for $\mathfrak{B}_{m}$ gravity coupled to spin-3 fields. Indeed, besides the
terms proportional to $\mu _{i+\bar{\imath}}$, which correspond to Euler
type CS\ term, there are CS terms related to the Pontryagin type densities,
which are proportional to the $\mu _{i+j}\,$\ and $\mu _{\bar{\imath}+\bar{j}%
}$ constants. The action (\ref{AccBk}) is invariant under $\mathfrak{B}%
_{m}^{s_{3}}$ gauge transformations with parameters of the form
\begin{equation}
\lambda =\Lambda ^{a,\left( i\right) }M_{a}^{\left( i\right) }+\xi
^{a,\left( \bar{\imath}\right) }P_{a}^{\left( \bar{\imath}\right) }+\Lambda
^{ab,\left( i\right) }M_{ab}^{\left( i\right) }+\xi ^{ab,\left( \bar{\imath}%
\right) }P_{ab}^{\left( \bar{\imath}\right) }\,,
\end{equation}%
which explicitly look like
\begin{eqnarray}
\delta \omega ^{a,\left( i\right) } &=&d\Lambda ^{a,\left( i\right) }+\delta
_{j+k}^{i}\epsilon ^{abc}\left( \omega _{b}^{\left( j\right) }\Lambda
_{c}^{\left( k\right) }-4\sigma \omega _{bd}^{\left( j\right) }\,\Lambda
_{c}^{\text{ }d,\left( k\right) }\right)   \notag \\
&&-\epsilon ^{abc}\left( \xi _{b}^{\left( \bar{j}\right) }e_{c}^{\left( \bar{%
k}\right) }+4\sigma e_{bd}^{\left( \bar{j}\right) }\xi _{c}^{\text{ }%
d,\left( \bar{k}\right) }\right) \delta _{\bar{j}+\bar{k}}^{i}\,, \label{tggen1}\\
\delta e^{a,\left( \bar{\imath}\right) } &=&d\xi ^{a,\left( \bar{\imath}%
\right) }+\delta _{j+\bar{k}}^{\bar{\imath}}\,\epsilon ^{abc}\left( \omega
_{b}^{\left( j\right) }\xi _{c}^{\left( \bar{k}\right) }-\Lambda
_{b}^{\left( j\right) }e_{c}^{\left( \bar{k}\right) }-4\sigma \omega
_{bd}^{\left( j\right) }\,\xi _{c}^{\text{ }d,\left( \bar{k}\right)
}-4\sigma e_{bd}^{\left( \bar{k}\right) }\,\Lambda _{c}^{\text{ }d,\left(
j\right) }\right)\, ,\label{tggen2}\\
\delta \omega ^{ab,\left( i\right) } &=&d\Lambda ^{ab,\left( i\right)
}+\delta _{j+k}^{i}\epsilon ^{cd\left( a\right\vert }\left( \omega
_{c}^{\left( j\right) }\Lambda _{d}^{\text{ }\left\vert b\right) ,\left(
k\right) }+\omega _{\text{ }c}^{b),\left( j\right) }\Lambda _{d}^{\left(
k\right) }\right)   \notag \\
&&+\delta _{\bar{j}+\bar{k}}^{i}\epsilon ^{cd\left( a\right\vert }\left(
e_{c}^{\left( \bar{j}\right) }\xi _{d}^{\text{ }\left\vert b\right) ,\left(
\bar{k}\right) }+e_{\text{ }c}^{b),\left( \bar{j}\right) }\xi _{d}^{\left(
\bar{k}\right) }\right) \,, \label{tggen3}\\
\delta e^{ab,\left( \bar{\imath}\right) } &=&d\xi ^{ab,\left( \bar{\imath}%
\right) }+\delta _{j+\bar{k}}^{\bar{\imath}}\epsilon ^{cd\left( a\right\vert
}\left( \omega _{c}^{\left( j\right) }\xi _{d}^{\text{ }\left\vert b\right)
,\left( \bar{k}\right) }+e_{c}^{\left( \bar{k}\right) }\Lambda _{d}^{\text{ }%
\left\vert b\right) ,\left( j\right) }+e_{\text{ }c}^{b),\left( \bar{k}%
\right) }\Lambda _{d}^{\left( j\right) }+\omega _{\text{ }c}^{b),\left(
j\right) }\xi _{d}^{\left( \bar{k}\right) }\right) \,.\label{tggen4}
\end{eqnarray}
The spin-3 gauge fields appear explicitly in the equation of
motion for the spin-2 fields, which are given by%
\begin{eqnarray}
&&\mathcal{T}^{a,\left( \bar{\imath}\right) }\equiv de^{a,\left( \bar{\imath}%
\right) }+\delta _{j+\bar{k}}^{\bar{\imath}}\left( \epsilon ^{abc}\omega
_{b}^{\left( j\right) }e_{c}^{\left( \bar{k}\right) }-4\sigma \epsilon
^{abc}e_{bd}^{\left( \bar{k}\right) }\omega _{c}^{\text{ }d,\left( j\right)
}\right) =0\,, \\
&&\mathcal{R}^{a,\left( i\right) }\equiv d\omega ^{a,\left( i\right) }+\frac{%
\delta _{j+k}^{i}}{2}\epsilon ^{abc}\left( \omega _{b}^{\left( j\right)
}\omega _{c}^{\left( k\right) }-4\sigma \omega _{bd}^{\left( j\right)
}\omega _{c}^{\text{ }d,\left( k\right) }\right)  \notag \\
&&\text{ \ \ }\text{\ \ \ \ \ \ \ \ \ \ }+\frac{\delta _{\bar{j}+\bar{k}}^{i}%
}{2}\epsilon ^{abc}\left( e_{b}^{\left( \bar{j}\right) }e_{c}^{\left( \bar{k}%
\right) }-4\sigma e_{bd}^{\left( \bar{j}\right) }e_{c}^{\text{ }d,\left(
\bar{k}\right) }\right) =0\,.
\end{eqnarray}%
On the other hand, the equations of motion for the spin-3 gauge fields read
\begin{eqnarray}
\mathcal{T}^{ab,\left( \bar{\imath}\right) } &\equiv &de^{ab,\left( \bar{%
\imath}\right) }+\delta _{j+\bar{k}}^{\bar{\imath}}\,\epsilon ^{cd\left(
a\right\vert }\left( \omega _{c}^{\left( j\right) }e_{d}^{\text{ }\left\vert
b\right) ,\left( \bar{k}\right) }+e_{c}^{\left( \bar{k}\right) }\omega _{d}^{%
\text{ }\left\vert b\right) ,\left( j\right) }\right) =0\,, \\
\mathcal{R}^{ab,\left( i\right) } &\equiv &d\omega ^{ab,\left( i\right)
}+\delta _{j+k}^{i}\epsilon ^{cd\left( a\right\vert }\left( \omega
_{c}^{\left( j\right) }\omega _{d}^{\text{ }\left\vert b\right) ,\left(
k\right) }\right) +\delta _{\bar{j}+\bar{k}}^{i}\epsilon ^{cd\left(
a\right\vert }e_{c}^{\left( \bar{j}\right) }e_{d}^{\text{ }\left\vert
b\right) ,\left( \bar{k}\right) }=0\,.
\end{eqnarray}%
Here the delta $\delta _{p+q}^{i}$ is restricted to $p+q\leq m-2$ due to the
non-vanishing components of the invariant tensor for the $\mathfrak{B}%
_{m}^{s_{3}}$ symmetry. It is interesting that, as in the HS Maxwell case, the field equations of the HS fields resemble the form of the ones of in \cite{FN, FNN} and could be a way to realize that kind of matter couplings from an algebraic point of view.

\subsection{$\mathfrak{C}_{m}$ gravity coupled to spin-3 fields}

The same procedure can be applied in order to obtain a $\mathfrak{C}_{m}$
algebra coupled to spin-3 generators. Indeed, let us consider now the semigroup $S_{%
\mathcal{M}}^{\left( m-2\right) }=\left\{ \lambda _{0},\lambda _{1},\dots
,\lambda _{m-2}\right\} $, whose elements
satisfy
\begin{equation}
\lambda _{\alpha }\lambda _{\beta }=\left\{
\begin{array}{lcl}
\lambda _{\alpha +\beta }\,\,\,\, & \mathrm{if}\,\,\,\,\alpha +\beta \leq m-2
&  \\
\lambda _{\alpha +\beta -2\left[ \frac{m-1}{2}\right] }\,\,\, & \mathrm{if}%
\,\,\,\,\alpha +\beta >m-2 &
\end{array}%
\right. .  \label{smk2}
\end{equation}%
The $S_{\mathcal{M}}^{\left( m-2\right) }$-expanded algebra has the same number
of generators as the $\mathfrak{B}_{m}^{s_{3}}$ algebra,
\begin{equation*}
\mathfrak{C}_{m}^{s_{3}}=S_{\mathcal{M}}^{\left( m-2\right) }\times
\mathfrak{sl}(3,\mathbb{R})=\left\{ M_{a}^{\left( i\right) },P_{a}^{\left( \bar{\imath}%
\right) },M_{ab}^{\left( i\right) },P_{ab}^{\left( \bar{\imath}\right)
}\right\} \,,
\end{equation*}%
and are related to the original ones through%
\begin{eqnarray}
\ell ^{i}M_{a}^{\left( i\right) } &=&J_{\left( a,i\right) }=\lambda
_{i}J_{a}\,,\text{ \ \ }\ell ^{i}M_{ab}^{\left( i\right) }=T_{\left(
ab,i\right) }=\lambda _{i}T_{ab}\,, \\
\ell ^{\bar{\imath}}P_{a}^{\left( \bar{\imath}\right) } &=&J_{\left( a,\bar{%
\imath}\right) }=\lambda _{\bar{\imath}}J_{a}\,,\text{ \ \ }\ell ^{\bar{%
\imath}}P_{ab}^{\left( \bar{\imath}\right) }=T_{\left( ab,\bar{\imath}%
\right) }=\lambda _{\bar{\imath}}T_{ab}\,.
\end{eqnarray}%
As in the previous case, $i$ takes even values, while $\bar{\imath}$
takes odd values. However, the $S_{\mathcal{M}}$ semigroup has no zero element,
 implying that the $\mathfrak{C}_{m}^{s_{3}}$ algebra has the form
\begin{eqnarray}
\left[ M_{a}^{\left( i\right) },M_{b}^{\left( j\right) }\right]  &=&\frac{%
\ell ^{\left\{ i+j\right\} }}{\ell ^{i+j}}\epsilon _{abc}M^{c,\left\{
i+j\right\} },\,\ \left[ M_{a}^{\left( i\right) },P_{b}^{\left( \bar{\imath}%
\right) }\right] =\frac{\ell ^{\left\{ i+\bar{\imath}\right\} }}{\ell ^{i+%
\bar{\imath}}}\epsilon _{abc}P^{c,\left\{ i+\bar{\imath}\right\} },\text{ \ }
\\
\left[ P_{a}^{\left( \bar{\imath}\right) },P_{b}^{\left( \bar{j}\right) }%
\right]  &=&\frac{\ell ^{\left\{ \bar{\imath}+\bar{j}\right\} }}{\ell ^{\bar{%
\imath}+\bar{j}}}\epsilon _{abc}M^{c,\left\{ \bar{\imath}+\bar{j}\right\}
},\,
\end{eqnarray}%
\begin{eqnarray}
\left[ M_{a}^{\left( i\right) },M_{bc}^{\left( j\right) }\right]  &=&\frac{%
\ell ^{\left\{ i+j\right\} }}{\ell ^{i+j}}\epsilon _{\text{ \ }a\left(
b\right. }^{m}M_{\left. c\right) m}^{\left\{ i+j\right\} }\,,\text{ \ \ }%
\left[ M_{a}^{\left( i\right) },P_{bc}^{\left( \bar{\imath}\right) }\right] =%
\frac{\ell ^{\left\{ i+\bar{\imath}\right\} }}{\ell ^{i+\bar{\imath}}}%
\epsilon _{\text{ \ }a\left( b\right. }^{m}P_{\left. c\right) m}^{\left\{ i+%
\bar{\imath}\right\} }\,, \\
\left[ P_{a}^{\left( \bar{\imath}\right) },M_{bc}^{\left( j\right) }\right]
&=&\frac{\ell ^{\left\{ \bar{\imath}+j\right\} }}{\ell ^{\bar{\imath}+j}}%
\epsilon _{\text{ \ }a\left( b\right. }^{m}P_{\left. c\right) m}^{\left\{
\bar{\imath}+j\right\} }\,,\text{ \ \ }\left[ P_{a}^{\left( \bar{\imath}%
\right) },P_{bc}^{\left( \bar{j}\right) }\right] =\frac{\ell ^{\left\{ \bar{%
\imath}+\bar{j}\right\} }}{\ell ^{\bar{\imath}+\bar{j}}}\epsilon _{\text{ \ }%
a\left( b\right. }^{m}M_{\left. c\right) m}^{\left\{ \bar{\imath}+\bar{j}%
\right\} }\,,
\end{eqnarray}%
\begin{eqnarray}
\left[ M_{ab}^{\left( i\right) },M_{cd}^{\left( j\right) }\right]  &=&\frac{%
\ell ^{\left\{ i+j\right\} }}{\ell ^{i+j}}\sigma \left( \eta _{a\left(
c\right. }\epsilon _{\left. d\right) bm}+\eta _{b\left( c\right. }\epsilon
_{\left. d\right) am}\right) M^{m,\left\{ i+j\right\} }, \\
\left[ M_{ab}^{\left( i\right) },P_{cd}^{\left( \bar{\imath}\right) }\right]
&=&\frac{\ell ^{\left\{ i+\bar{\imath}\right\} }}{\ell ^{i+\bar{\imath}}}%
\sigma \left( \eta _{a\left( c\right. }\epsilon _{\left. d\right) bm}+\eta
_{b\left( c\right. }\epsilon _{\left. d\right) am}\right) P^{m,\left\{ i+%
\bar{\imath}\right\} }, \\
\left[ P_{ab}^{\left( i\right) },P_{cd}^{\left( \bar{j}\right) }\right]  &=&%
\frac{\ell ^{\left\{ \bar{\imath}+\bar{j}\right\} }}{\ell ^{\bar{\imath}+%
\bar{j}}}\sigma \left( \eta _{a\left( c\right. }\epsilon _{\left. d\right)
bm}+\eta _{b\left( c\right. }\epsilon _{\left. d\right) am}\right)
M^{m,\left\{ \bar{\imath}+\bar{j}\right\} },
\end{eqnarray}%
where $\left\{ \cdots \right\} $ means%
\begin{equation}
\left\{ i+j\right\} =\left\{
\begin{array}{lcl}
i+j & \mathrm{if}\,\,\,\,i+j\leq m-2 &  \\
i+j-2\left[ \frac{m-1}{2}\right] \, & \mathrm{if}\,\,\,\,i+j>m-2 &
\end{array}%
\right. .
\end{equation}

The $\mathfrak{C}_{m}^{s_{3}}$ algebra describes the coupling of spin-3
generators $\left\{ M_{ab}^{\left( i\right) },P_{ab}^{\left( \bar{\imath}%
\right) }\right\} $ to the spin-2 generators $\left\{ M_{a}^{\left( i\right)
},P_{a}^{\left( \bar{\imath}\right) }\right\} $ of the $\mathfrak{C}_{m}$
Lie algebra. Let us note that the $\mathfrak{sl}\left( 3,
\mathbb{R}
\right) \times \mathfrak{sl}\left( 3,
\mathbb{R}
\right) $ algebra (\ref{2sl3A}-\ref{2sl3B}) is recovered for $m=3$, and $m=4$ reproduces
the $AdS$-Lorentz Lie algebra coupled to spin-3 generators, which in turn can be rewritten as three copies of the
$\mathfrak{sl}\left( 3,
\mathbb{R}
\right) $ algebra. This fact might lead one to think that the $\mathfrak{C}%
_{m}^{s_{3}}$ algebra could be isomorphic to $m-1$ copies of $\mathfrak{sl}%
\left( 3,%
\mathbb{R}
\right) $. However, this is an accident that occurs only for $m=3,4$, and it is not
true in the generic case. The $\mathfrak{C}_{m}^{s_{3}}$ algebra reduces to the $\mathfrak{B}_{m}^{s_{3}}$ alebra described in the previous section when the
limit $\ell \rightarrow \infty $ is considered.  In such limit, $\ell ^{\left\{
p+q\right\} }/\ell ^{p+q}\rightarrow 0$ for $p+q>m-2$, which abelianizes some
commutation relations.
The quadratic
Casimir in this case reads
\begin{eqnarray}
C &=&\frac{\mu _{\left\{ i+j\right\} }}{\ell ^{i+j}}\left(
\sum_{i,j}^{m-2}M_{a}^{\left( i\right) }M^{a,\left( j\right) }-\frac{1}{%
2\sigma }M_{ab}^{\left( i\right) }M^{ab,\left( j\right) }\right)
+\frac{\mu _{\left\{ i+\bar{\imath}\right\} }}{\ell ^{i+\bar{\imath}}}%
\left( \sum_{i,\bar{\imath}}^{m-2}M_{a}^{\left( i\right) }P^{a,\left( \bar{%
\imath}\right) }-\frac{1}{2\sigma }M_{ab}^{\left( i\right) }P ^{ab,\left(%
\bar{\imath}\right) }\right)   \notag \\
&&+\frac{\mu _{\left\{ \bar{\imath}+\bar{j}\right\} }}{\ell ^{\bar{\imath}+%
\bar{j}}}\left( \sum_{\bar{\imath},\bar{j}}^{m-2}\,P_{a}^{\left( \bar{\imath}%
\right) }P^{a,\left( \bar{j}\right) }-\frac{1}{2\sigma }P_{ab}^{\left( \bar{%
\imath}\right) }P^{ab,\left( \bar{j}\right) }\right) \,.
\end{eqnarray}%
The limit $\ell \rightarrow \infty $ can also be applied at
the level of the $\mathfrak{C}_{m}^{s_{3}}$ Casimir operator.
Nevertheless, this requires the following redefinition in the arbitrary
constants,
\begin{equation}
\mu _{\left\{ i+j\right\} }\rightarrow \ell ^{\left\{ i+j\right\} }\mu
_{\left\{ i+j\right\} }\,.  \label{redefconst}
\end{equation}%
Then, as in the commutation relations, $\ell ^{\left\{ p+q\right\}
}/\ell ^{p+q}$ vanishes for $p+q>m-2$, which reproduces the $\mathfrak{B}%
_{m}^{s_{3}}$ quadratic Casimir.

Let us consider the $\mathfrak{C}_{m}^{s_{3}}$ connection one-form
\begin{equation}
A=\omega ^{a,\left( i\right) }M_{a}^{\left( i\right) }+e^{a,\left( \bar{%
\imath}\right) }P_{a}^{\left( \bar{\imath}\right) }+\omega ^{ab,\left(
i\right) }M_{ab}^{\left( i\right) }+e^{ab,\left( \bar{\imath}\right)
}P_{ab}^{\left( \bar{\imath}\right) }\,.  \label{ofck}
\end{equation}%
and non-vanishing
components of the invariant tensor for the $\mathfrak{C}_{m}^{s_{3}}$ algebra
\begin{eqnarray}
\left\langle M_{a}^{\left( i\right) }M_{b}^{\left( j\right) }\right\rangle
&=&\frac{\mu _{\left\{ i+j\right\} }}{2\ell ^{i+j}}\,\eta _{ab}\,,\text{ \ \
}\left\langle M_{a}^{\left( i\right) }P_{b}^{\left( \bar{\imath}\right)
}\right\rangle =\frac{\mu _{\left\{ i+\bar{\imath}\right\} }}{2\ell ^{i+\bar{%
\imath}}}\,\eta _{ab}\,,\text{ \ \ }\left\langle P_{a}^{\left( \bar{\imath}%
\right) }P_{b}^{\left( \bar{j}\right) }\right\rangle =\frac{\mu _{\left\{
\bar{\imath}+\bar{j}\right\} }}{2\ell ^{\bar{\imath}+\bar{j}}}\,\eta _{ab}\,,
\\
\left\langle M_{ab}^{\left( i\right) }M_{cd}^{\left( j\right) }\right\rangle
&=&-\frac{\sigma }{2\ell ^{i+j}}\mu _{\left\{ i+j\right\} }\left( \eta
_{a\left( c\right. }\epsilon _{\left. d\right) b}-\frac{2}{3}\eta _{ab}\eta
_{cd}\right) \,, \\
\left\langle M_{ab}^{\left( i\right) }P_{cd}^{\left( \bar{\imath}\right)
}\right\rangle &=&-\frac{\sigma }{2\ell ^{i+\bar{\imath}}}\mu _{\left\{ i+%
\bar{\imath}\right\} }\left( \eta _{a\left( c\right. }\epsilon _{\left.
d\right) b}-\frac{2}{3}\eta _{ab}\eta _{cd}\right) \,, \\
\left\langle P_{ab}^{\left( \bar{\imath}\right) }P_{cd}^{\left( \bar{j}%
\right) }\right\rangle &=&-\frac{\sigma }{2\ell ^{\bar{\imath}+\bar{j}}}\mu
_{\left\{ \bar{\imath}+\bar{j}\right\} }\left( \eta _{a\left( c\right.
}\epsilon _{\left. d\right) b}-\frac{2}{3}\eta _{ab}\eta _{cd}\right) \,,
\end{eqnarray}%
where $\left\{ \cdots \right\} $ is defined as%
\begin{equation}
\left\{ i+j\right\} =\left\{
\begin{array}{lcl}
i+j & \mathrm{if}\,\,\,\,i+j\leq m-2 &  \\
i+j-2\left[ \frac{m-1}{2}\right] \, & \mathrm{if}\,\,\,\,i+j>m-2 &
\end{array}%
\right. .
\end{equation}%
The corresponding three-dimensional CS action (\ref{eq:action}) is
\begin{eqnarray}
S &=&\kappa\int \frac{\mu _{\left\{ i+j\right\} }}{\ell ^{i+j}}\,%
\left[ \frac{1}{2}\left( \omega ^{a,\left( i\right) }d\omega _{a}^{\left(
j\right) }+\frac{\ell ^{\left\{ i+j\right\} }}{3\ell ^{i+j}}\epsilon
_{abc}\omega ^{a,\left( l\right) }\omega ^{b,\left( m\right) }\omega
^{c,\left( n\right) }\,\delta _{l+m+n}^{\left\{ i+j\right\} }\right) \right.
\notag \\
&&\left. -\sigma \left( \omega _{\text{ }b}^{a,\left( i\right) }d\omega _{%
\text{ }a}^{b,\left( j\right) }+\frac{2\ell ^{\left\{ i+j\right\} }}{\ell
^{i+j}}\epsilon _{abc}\omega ^{a,\left( l\right) }\omega ^{bd,\left(
m\right) }\omega _{\text{ }d}^{c,\left( n\right) }\,\delta _{l+m+n}^{\left\{
i+j\right\} }\right) \right]   \notag \\
&&+\frac{\mu _{\left\{ i+\bar{\imath}\right\} }}{\ell ^{i+\bar{\imath}}}%
\left[ e^{a,\left( \bar{\imath}\right) }\left( d\omega _{a}^{\left( i\right)
}+\frac{\ell ^{\left\{ i+\bar{\imath}\right\} }}{2\ell ^{i+\bar{\imath}}}%
\epsilon _{abc}\omega ^{b,\left( m\right) }\omega ^{c,\left( n\right)
}\,\delta _{m+n}^{i}-\frac{2\ell ^{\left\{ i+\bar{\imath}\right\} }}{\ell
^{i+\bar{\imath}}}\sigma \epsilon _{abc}\omega ^{bd,\left( m\right) }\omega
_{\text{ }d}^{c,\left( n\right) }\delta _{m+n}^{i}\right) \right.   \notag \\
&&\left. -2\sigma e^{ab,\left( \bar{\imath}\right) }\left( d\omega _{\text{ }%
a}^{b,\left( i\right) }+\frac{2\ell ^{\left\{ i+\bar{\imath}\right\} }}{\ell
^{i+\bar{\imath}}}\epsilon _{acd}\omega ^{c,\left( m\right) }\omega _{\text{
}b}^{d,\left( n\right) }\,\delta _{m+n}^{i}\right) \right]   \notag \\
&&+\frac{\mu _{\left\{ \bar{\imath}+\bar{j}\right\} }}{\ell ^{\bar{\imath}+%
\bar{j}}}\left[ \frac{1}{2}e^{a,\left( \bar{\imath}\right) }\left(
de_{a}^{\left( \bar{j}\right) }+\frac{\ell ^{\left\{ \bar{\imath}+\bar{j}%
\right\} }}{\ell ^{\bar{\imath}+\bar{j}}}\epsilon _{abc}\omega ^{b,\left(
m\right) }e^{c,\left( \bar{n}\right) }\delta _{m+\bar{n}}^{\bar{j}}\right)
\right.   \notag \\
&&\left. -\sigma e_{\text{ }}^{ab,\left( \bar{\imath}\right) }\left(
de_{ab}^{\left( \bar{j}\right) }+\frac{2\ell ^{\left\{ \bar{\imath}+\bar{j}%
\right\} }}{\ell ^{\bar{\imath}+\bar{j}}}\epsilon _{acd}\omega ^{c,\left(
m\right) }e_{\text{ }b}^{d,\left( \bar{n}\right) }\delta _{m+\bar{n}}^{\bar{j%
}}+\frac{4\ell ^{\left\{ \bar{\imath}+\bar{j}\right\} }}{\ell ^{\bar{\imath}+%
\bar{j}}}\epsilon _{acd}e^{c,\left( \bar{m}\right) }\omega _{\text{ }%
b}^{d,\left( n\right) }\delta _{\bar{m}+n}^{\bar{j}}\right) \right] \,.
\label{csck}
\end{eqnarray}%
This action describes a spin-3 extension of $\mathfrak{C}%
_{m}$ gravity and is divided in two parts. The term
proportional to $\mu _{\left\{ i+\bar{\imath}\right\} }$ corresponds to an
Euler type CS action while the terms proportional to $\mu _{\left\{
i+j\right\} }$ and $\mu _{\left\{ \bar{\imath}+\bar{j}\right\} }$ describe\
a Pontryagin type CS action.
Note that $\ell ^{\left\{ p+q\right\} }/\ell ^{p+q}$ is trivially the
identity for $p+q\leq m-2$. Then the parameter $\ell $ appears in the
action for $p+q>m-2$. In orther to apply the limit $\ell
\rightarrow \infty $ in the CS action, it is necessary to use the redefinition (\ref{redefconst}%
), leading to the $\mathfrak{B}_{m}^{s_{3}}$ action (\ref{AccBk}).
The field equations coming from the action (\ref{csck}) are
 \begin{eqnarray}
&&\mathcal{T}^{a,\left( \bar{\imath}\right) }\equiv de^{a,\left( \bar{\imath}%
\right) }+\delta _{\left\{ j+\bar{k}\right\} }^{\bar{\imath}}\left( \epsilon
^{abc}\omega _{b}^{\left( j\right) }e_{c}^{\left( \bar{k}\right) }-4\sigma
\epsilon ^{abc}e_{bd}^{\left( \bar{k}\right) }\omega _{c}^{\text{ }d,\left(
j\right) }\right) =0\,, \\
&&\mathcal{R}^{a,\left( i\right) }\equiv d\omega ^{a,\left( i\right) }+\frac{%
\delta _{\left\{ j+k\right\} }^{i}}{2}\epsilon ^{abc}\left( \omega
_{b}^{\left( j\right) }\omega _{c}^{\left( k\right) }-4\sigma \omega
_{bd}^{\left( j\right) }\omega _{c}^{\text{ }d,\left( k\right) }\right)
\notag \\
&&\text{ \ \ \ \ \ \ \ \ \ \ \ \ }+\frac{\delta _{\bar{j}+\bar{k}}^{i}}{%
2\ell ^{2}}\epsilon ^{abc}\left( e_{b}^{\left( \bar{j}\right) }e_{c}^{\left(
\bar{k}\right) }-4\sigma e_{bd}^{\left( \bar{j}\right) }e_{c}^{\text{ }%
d,\left( \bar{k}\right) }\right) =0\,,
\end{eqnarray}%
\begin{eqnarray}
\mathcal{T}^{ab,\left( \bar{\imath}\right) } &\equiv &de^{ab,\left( \bar{%
\imath}\right) }+\delta _{\left\{ j+\bar{k}\right\} }^{\bar{\imath}%
}\,\epsilon ^{cd\left( a\right\vert }\left( \omega _{c}^{\left( j\right)
}e_{d}^{\text{ }\left\vert b\right) ,\left( \bar{k}\right) }+e_{c}^{\left(
\bar{k}\right) }\omega _{d}^{\text{ }\left\vert b\right) ,\left( j\right)
}\right) =0\,, \\
\mathcal{R}^{ab,\left( i\right) } &\equiv &d\omega ^{ab,\left( i\right)
}+\delta _{\left\{ j+k\right\} }^{i}\epsilon ^{cd\left( a\right\vert }\left(
\omega _{c}^{\left( j\right) }\omega _{d}^{\text{ }\left\vert b\right)
,\left( k\right) }\right) +\frac{\delta _{\bar{j}+\bar{k}}^{i}}{\ell ^{2}}%
\epsilon ^{cd\left( a\right\vert }e_{c}^{\left( \bar{j}\right) }e_{d}^{\text{
}\left\vert b\right) ,\left( \bar{k}\right) }=0\,.
\end{eqnarray}%
We see that, although the
action (\ref{csck}) is quite similar to the $\mathfrak{B}_{m}^{s_{3}}$ one,
the absence of abelian generators in the $\mathfrak{C}_{m}^{s_{3}}$ symmetry
radically modifies the gravity action and therefore its dynamics.
Regarding gauge transformations, they differ from the ones of the $\mathfrak{B}%
_{m}^{s_{3}}$ case, (\ref{tggen1}-\ref{tggen4}) , in the Kroenecker delta $\delta _{p+q}^{i}$, which is no
more restricted to particular values of $p$ and $q$.

\section{Conclusions}

We have constructed new spin-3 extensions of Einstein gravity in three dimensions. This has been achieved by expanding the $\mathfrak{sl}(3,\mathbb{R})$ algebra.
As a warm up, we first showed how the known spin-3 extensions of the the $AdS$ and Poincar\'e algebras can be obtained as expansions of $\mathfrak{sl}(3,\mathbb{R})$. Using this technique, we have constructed most general invariant tensor in each case  and constructed the corresponding CS actions. This reproduces the known HS gravity theories in each case plus exotic terms.

By considering more general semigroups we have constructed the spin-3 extensions of the Maxwell and $AdS$-Lorentz algebras. In addition, the CS actions invariant under these new HS symmetries have been studied. In the case of the Maxwell algebra, it allows us to introduce a new spin-3 theory with vanishing cosmological constant. This is a novel extension of HS three-dimensional gravity in flat space including topological HS matter. The field equations of the HS field resemble the form of the matter field equations introduced in \cite{FN, FNN}. Therefore the formulation here presented could be a way to realize that kind of matter couplings from an algebraic point of view. At the same time we have showed how this model can be obtained as a flat limit of the CS theory corresponding the spin-3 extension of the $AdS$-Lorentz algebra.  Moreover, the flat limit relating
such new spin-3 gravity theories works at the level of the symmetries, Casimir operators, invariant tensors and field equations.

Finally, we have  generalized these results to construct two families of spin-3 algebras, denoted as $\mathfrak{C}_{m}^{s_{3}}$ and $\mathfrak{B}_{m}^{s_{3}}$, which contain all the previously obtained results as particular cases. These symmetries are related through the limit $\ell \rightarrow \infty $ for every value of $m$. The corresponding CS theories have been studied in each case, yielding new gravity models coupled to HS topological matter in three dimensions.

It would be interesting to go further and analyze the asymptotic symmetries of
these new HS theories, as well as their classical duals. This could lead to enlarged versions of known $\mathcal{W}$-type algebras. In particular, in the Maxwell
spin-$3$ case, one could expect an enlargement of the HS extension of the $\mathfrak{bms}_{3}$ algebra (work in progress).
Another appealing aspect of these theories is their solution space. It would be interesting to study, for instance, black holes, cosmologies and conical singularities supporting HS spin matter and analyze their thermodynamics. On the other hand, in order to describe more general models, it would be
worth to extend our analysis to an infinite number of interacting
HS gauge fields. Another important aspect along this direction that deserves further
investigation is the generalization of our results to include fermions.

\section{Acknowledgment}

The authors would like to thank to X. Bekaert and A. Campoleoni for
enlightening discussions and comments. This work was supported by the
Chilean FONDECYT Projects N$^{\circ }$3170437 (P.C.), N$^{\circ }$3170438
(E.R.) and N$^{\circ }$3160581 (P.S-R.). R.C. and O.F. would like to thank
to the Direcci\'{o}n de Investigaci\'{o}n and Vice-rector\'{\i}a de
Investigaci\'{o}n of the Universidad Cat\'{o}lica de la Sant\'{\i}sima
Concepci\'{o}n, Chile, for their constant support.

\end{document}